\documentclass[journal]{IEEEtran}
\usepackage{graphicx} 
\usepackage{subfigure}
\usepackage{amsmath}
\usepackage{multirow} 
\newtheorem{definition}{\textbf {Definition}}

\usepackage{caption}

\newcommand{\re}[1]{{\color[rgb]{0,0,0}#1}}

\usepackage{algorithm}
\usepackage{algorithmic}

\usepackage{amsfonts}
\usepackage{algorithm, algorithmic}
\usepackage{color}
\usepackage{booktabs}
\usepackage{array}
\usepackage{epstopdf} 
\usepackage{url}
\usepackage{cite}

\begin{document}

\title{Enabling Deep Learning-based Physical-layer Secret Key Generation for FDD-OFDM Systems in Multi-Environments}

\author{Xinwei~Zhang, 
        Guyue~Li,~\IEEEmembership{Member,~IEEE},
        Junqing~Zhang,~\IEEEmembership{Member,~IEEE},
        Linning~Peng,~\IEEEmembership{Member,~IEEE},
        Aiqun~Hu,~\IEEEmembership{Senior Member,~IEEE}
        and Xianbin~Wang,~\IEEEmembership{Fellow,~IEEE}
\thanks{
Copyright (c) 2015 IEEE. Personal use of this material is permitted. However, permission to use this material for any other purposes must be obtained from the IEEE by sending a request to pubs-permissions@ieee.org. 

Manuscript received 20 May 2023; revised 8 November 2023; accepted 8 February 2024.
This work was supported in part by the National Key R\&D Program of China (No. 2022YFB2902202), in part by the National Natural Science Foundation of China (No. 62171121, U22A2001) and in part by the National Natural Science Foundation of Jiangsu Province, China (No. BK20211160). 
The work of J. Zhang was in part supported by the UK Engineering and Physical Sciences Research Council (EPSRC) New Investigator Award under grant ID EP/V027697/1. 
The review of this paper was coordinated by Dr. Linke Guo. 
(\textit{Corresponding author: Guyue Li})}				
\IEEEcompsocitemizethanks{
\IEEEcompsocthanksitem Xinwei Zhang, Guyue Li, and Linning Peng are with the School of Cyber Science and Engineering, Southeast University, Nanjing 210096, China (e-mail: xwzhang1998@gmail.com, guyuelee@seu.edu.cn, pengln@seu.edu.cn).
\IEEEcompsocthanksitem Junqing Zhang is with the Department of Electrical Engineering and Electronics, University of Liverpool, Liverpool L69 3GJ, U.K. (e-mail: junqing.zhang@liverpool.ac.uk).
\IEEEcompsocthanksitem Aiqun Hu is with the School of Information Science and Engineering, and National Mobile Communications Research Laboratory, Southeast University, Nanjing 210096, China (e-mail: aqhu@seu.edu.cn).
\IEEEcompsocthanksitem Guyue Li, Linning Peng and Aiqun Hu are also with the Purple Mountain Laboratories for Network and Communication Security, Nanjing 210096, China.
\IEEEcompsocthanksitem Xianbin Wang is with the Department of Electrical and Computer Engineering, Western University, London, ON N6A 5B9, Canada (e-mail: xianbin.wang@uwo.ca).}
}

\markboth{IEEE Transactions on Vehicular Technology,~Vol.~XX, No.~XX, XXX~}
{}

\maketitle

\begin{abstract}
Deep learning-based physical-layer secret key generation (PKG) has been used to overcome the imperfect uplink/downlink channel reciprocity in frequency division duplexing (FDD) orthogonal frequency division multiplexing (OFDM) systems. However, existing efforts have focused on key generation for users in a specific environment where the training samples and test samples follow the same distribution, which is unrealistic for real-world applications. This paper formulates the PKG problem in multiple environments as a learning-based problem by learning the knowledge such as data and models from known environments to generate keys quickly and efficiently in multiple new environments. Specifically, we propose deep transfer learning (DTL) and meta-learning-based channel feature mapping algorithms for key generation. The two algorithms use different training methods to pre-train the model in the known environments, and then quickly adapt and deploy the model to new environments. Simulation and experimental results show that compared with the methods without adaptation, the DTL and meta-learning algorithms both can improve the performance of generated keys. In addition, the complexity analysis shows that the meta-learning algorithm can achieve better performance than the DTL algorithm with less cost.
\end{abstract}



\section{Introduction}

\IEEEPARstart{D}{ue} to the broadcast nature of radio signal propagation, wireless networks are vulnerable to various attacks such as eavesdropping, impersonating, and tampering~\cite{zou2016survey}. Traditional security mechanisms, particularly public key cryptography, are facing many problems such as difficulty in key distribution and poor scalability in large-scale networks with limited resources, which make it difficult to meet the security needs of future wireless communications \cite{li2019physical}. In recent years, \textit{physical-layer secret key generation (PKG)} has gradually become a research hotspot of wireless security. From the perspective of information theory, PKG provides a new security mechanism, which greatly simplifies the distribution and management of keys \cite{li2019physical, 8735939}. 

PKG techniques realize real-time sharing and coordination of random security keys by exploiting the channel reciprocity of uplink and downlink features~\cite{li2019physical, li2018constructing}. \re{Channel reciprocity means that both communicating parties can obtain highly similar channel characteristics, which determines whether the communicating parties can generate consistent keys. } Different channel features, such as received signal strength (RSS), channel state information (CSI), channel gain, etc., are widely used for PKG \cite{li2019physical}. In time division duplexing (TDD) systems, as both the uplink and downlink transmissions operate in the same carrier frequency band, and the channel features observed by both communication parties are highly reciprocal. However, in frequency division duplexing (FDD) systems, since the uplink and downlink are operated in different bands, the channel parameters observed by the two parties involved may be completely different, thus can not be directly used for key generation. Therefore, the majority of the existing studies focus on PKG in TDD systems and the research on PKG in FDD systems is limited. However, it has profound research value and practical significance to develop effective PKG solutions for FDD systems since FDD is the primary duplexing technique for cellular communications\cite{penttinen2015telecommunications, 3gpp.38.101-1}.

\subsection{Related Work}
In recent years, there have been some studies on the PKG in FDD systems, which can be categorized into model-based and deep learning-based approaches. 

Model-based methods aim to extract frequency-independent channel features or construct a reciprocal feature. Specifically, the work in~\cite{wang2012wireless,liu2019secret} proposed to extract the frequency-independent channel parameters (such as arriving angle, delay and covariance matrix eigenvalues). However, these methods have many limitations, such as large bandwidth or special configuration of the antenna array \cite{vasisht2016eliminating}. Besides, A framework for constructing reciprocal channels via path separation, adjustment and reconstruction is proposed in \cite{li2018constructing}, but it is difficult to separate the channel paths accurately in the complex multi-path environment. Some works proposed to construct reciprocal channels by additional reverse channel training and feedback, called the loopback-based methods~\cite{qin2016exploiting,allam2017channel}. However, these methods not only increase the complexity of channel detection but also increase the possibility of eavesdropping~\cite {linning2018investigation}. 

Due to its excellent performance, deep learning has also been introduced into the field of PKG in FDD systems \cite{zhang2021secret,wan2021secret,10.1145/3522783.3529526,hou2021secret,zhang2022deep}. In FDD systems, deep learning-based approaches have been used to construct reciprocal features for key generation with the help of the feature mapping function between uplink and downlink transmissions assisted by deep learning.
Since the uplink and downlink channels pass through the same propagation path and scattering clusters, it is experimentally shown in \cite{vasisht2016eliminating} that there is a transformation function that can map the channel to the underlying path. Furthermore, prior works have shown that it is possible to infer downlink channels from uplink channels~\cite{alrabeian2019deep,bakshi2019fast,liu2021Fire}. 
These works inspire efforts to apply deep learning for FDD-based key generation by constructing reciprocal features via deep learning.
In \cite{zhang2022deep}, it is proved that in a given environment, when the channel mapping function of possible user locations to antennas is bijective, there exists a feature mapping function that can map one frequency band features to another frequency band features, and the channel feature mapping function can be obtained by a simple deep learning model. This conclusion provides a theoretical basis for introducing deep learning into key generation for FDD-based OFDM systems~\cite{zhang2021secret,wan2021secret,hou2021secret,zhang2022deep}. A boundary equilibrium generative
adversarial network (BEGAN) and an encoder-decoder-based convolutional neural network were proposed to predict downlink CSI and key generation \cite{wan2021secret,hou2021secret}. Furthermore, a complex-valued neural network (CVNet) was proposed to improve the performance of generated keys \cite{zhang2021secret}. 

Compared with conventional model-based PKG techniques (e.g. \cite{wang2012wireless,liu2019secret,qin2016exploiting,allam2017channel,li2018constructing}), deep learning-based key generation methods are not limited to channel models and can achieve excellent performance. However, existing deep learning-based approaches only consider a given wireless environment and the deep learning model only can learn the feature mapping function in this specific environment. In practice, users may experience different new environments. 
Existing machine learning techniques require data collection and model training for each communication environment, leading to a large number of training resources and training data, which is difficult to be used in real-world applications. Therefore, how to quickly adapt the deep learning model to new environments for feature mapping and key generation with low cost is a new challenge that needs to be addressed.

Deep transfer learning (DTL)~\cite{nguyen2021transfer} and meta-learning \cite{thrun1998learning,park2021learning} are effective ways that can solve the problem of inapplicability of the deep learning model caused by environmental changes. DTL uses the knowledge of source tasks to improve the performance of target tasks and is a promising machine learning technology that can solve similar tasks with limited labeled data. Meta-learning aims to improve the ability to adapt or generalize to new tasks and environments that have never been encountered during the training stage by training in multiple learning tasks.
They have been widely used in many areas to solve the problem of performance degradation of deep learning models due to environmental changes, e.g., channel feedback \cite{zeng2021downlink}, beam prediction \cite{yuan2020transfer}, downlink channel prediction \cite{yang2020deep}, 
Beamforming Optimization \cite{9367008} etc, but still not applied to the field of PKG.

\subsection{Main Contributions}
\re{Inspired by these works, this paper introduces DTL and meta-learning into the field of PKG to improve channel reciprocity for achieving the fast and efficient key generation of FDD-OFDM systems in multi-environments.} \re{In our work, we focus on generating high-performance keys and adopting optimal neural network architectures and training strategies.}
First, we formulate the key generation in multi-environments as a learning-based problem, i.e., using the knowledge from known (source) environments to learn the deep learning model in the new (target) environments more efficiently. Then we propose DTL-based and meta-learning-based feature mapping algorithms to achieve key generation for FDD systems in multi-environments and verify the performance of the proposed algorithm with sufficient simulation and experimental data. 
\re{To the best knowledge of the authors, this is the first work focusing on deep learning-based key generation for FDD-OFDM systems in multi-environments.} Our major contributions are summarized as follows.
\begin{itemize}
	\item We propose a DTL-based channel feature mapping algorithm for physical layer key generation in FDD-OFDM systems. This algorithm pre-trains the model using the datasets from source environments and then fine-tunes the pre-trained model using a small number of samples from the new environment, after which this fine-tuned model can be used for key generation in the new environment. \re{The algorithm can be fine-tuned on the model of the current systems without needing a pre-training process, allowing for seamless integration into the existing system.}
	\item To better leverage knowledge from known channel environments, we propose a meta-learning-based feature mapping algorithm for key generation in FDD-OFDM systems. This algorithm performs intra-task and cross-task learning in multiple tasks (each task represents the key generation in a given environment) to obtain the best model initialization parameters, allowing for fast model adaptation in new environments.
    \item \re{We analyze channel differences in different environments, and give a theoretical analysis of the DTL and meta-learning algorithms. The results show that the data distribution in different environments is almost different, which is the cause of the better performance of the meta-learning algorithm.}
	\item We verify the proposed algorithms in an outdoor scenario using a ray tracing simulator Wireless InSite. The results show the DTL and meta-learning algorithms can both improve the performance of generated keys in new environments. In addition, complexity analysis shows that the meta-learning algorithm can achieve better performance with less time cost compared with the DTL algorithm. 
 
    \item A practical GNURadio and USRP-based FDD-OFDM wireless key generation prototype is developed. We collect data in both indoor and outdoor scenarios and for the first time validate the performance of the deep learning-powered FDD-OFDM key generation method using real-world data. We further verified through real experiments that the proposed algorithm can significantly reduce the key error rate (KER) of the key generated in the new environment. \re{There are currently no works in related fields that use data in real-life environments for verification. }
\end{itemize} 

The rest of this paper is structured as follows. The deep learning-based key generation for FDD-OFDM systems is introduced in Section \ref{Preliminary}. In Section \ref{system overview}, we formulate the PKG in multi-environments as a learning-based problem and give an algorithm overview. The DTL and meta-learning-based feature mapping algorithms for key generation are presented in Section \ref{DTL} and Section \ref{meta}. The DTL and meta-learning are analyzed theoretically in Section \ref{theoretical}. The simulation results for evaluating the performance of the generated keys and the complexity analysis are provided in Section \ref{simulation}. In Section \ref{practical}, we built a platform and verified the performance of the algorithm in a real environment, which is followed by conclusions in Section \ref{Conclusion}.

\section{Preliminary: Deep Learning-Powered FDD-OFDM Key Generation}
\label{Preliminary}

\subsection{Overview}
We consider the FDD-OFDM system, where the BS (Alice) and user (Bob) are equipped with a single antenna and operate in the FDD mode. Alice and Bob simultaneously transmit signals on different carrier frequencies, $f_{dl}$ and $f_{ul}$, respectively. The channel impulse response (CIR) can be defined as follows:
\begin{equation}
\begin{split}
h(f,\tau)=\sum_{n=0}^{N-1}\alpha_{n}e^{-j2\pi f\tau_n+j\phi_n}\delta(\tau-\tau_n),
\end{split}
\label{CIR}
\end{equation}
where $f$ is the carrier frequency, $N$ is the total number of paths, $\alpha_{n}$ is the magnitude of the $n^{th}$ path, which is influenced by the distance $d_n$ between Alice and Bob, the scattering environment and the carrier frequency  $f$. $\tau_n = \frac{d_n}{c}$ is the delay of the $n^{th}$ path, where $c$ is the speed of light. $\phi_n$ is the phase shift of the $n^{th}$ path, which is determined by the scatterer material and wave incident/impinging angles at the scatterer.


In FDD-OFDM systems, the channel frequency response (CFR) of the $l^{th}$ sub-carrier can be expressed as 
\begin{equation}
\begin{split}
H(f,l)=\sum_{n=0}^{N-1}\alpha_{n}e^{-j2\pi f\tau_n+j\phi_n}e^{-j2\pi n f_l},
\end{split}
\label{CFR}
\end{equation}
where $f_l$ is the frequency of the $l^{th}$ subcarrier relative to the
center frequency $f$. The CFR of frequency $f$ can be defined as the $1\times L$ channel vector ${\mathbf{H}(f)}=\{H(f,0),...,H(f,L-1)\} $, where $L$ is the total number of sub-carrier.
As shown in (\ref{CFR}), the amplitude and phase of wireless channel ${\mathbf{H}(f)}$ are influenced by their frequencies. 
Therefore, extracting reciprocal channel features for key generation in FDD-OFDM systems is challenging.

\begin{figure}[!t]
	\centering
	\includegraphics[width=\linewidth]{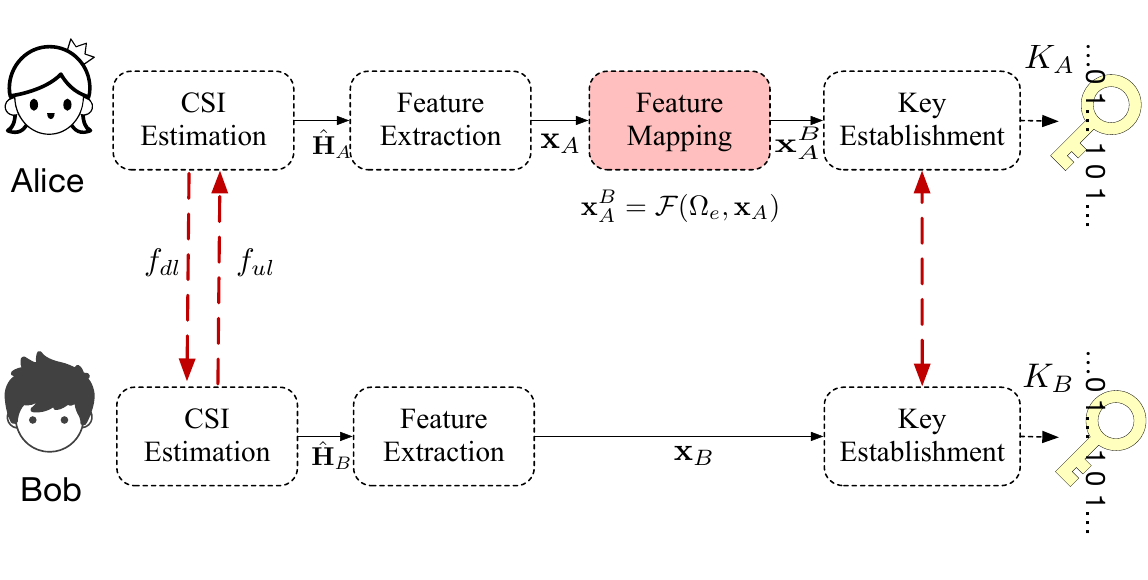}
	\caption{Deep learning-based key generation for FDD-OFDM systems.}
	\label{DL_KG}
\end{figure}

Deep learning has been introduced for PKG in FDD-OFDM systems recently \cite{zhang2021secret,wan2021secret,hou2021secret,zhang2022deep}. This type of method uses deep learning techniques to map the uplink features to the downlink features, so that both parties can obtain the downlink features at the same time. As shown in Fig. \ref{DL_KG}, the deep learning-based key generation contains the following four steps.




\subsection{CSI Estimation}
Alice and Bob simultaneously send OFDM pilot signals to each other at carrier frequencies $f_{dl}$ and $f_{ul}$, and then independently estimate the channel CFR based on the received pilot signals, expressed as 
\begin{equation}
	\begin{split}
		\begin{cases}
			\hat{H}_{A}\left(f_{ul}, l\right)=H\left(f_{ul}, l\right)+E_{1}\left(f_{ul}, l\right) \\
			\hat{H}_{B}\left(f_{dl}, l\right)=H\left(f_{dl}, l\right)+E_{2}\left(f_{dl}, l\right)
		\end{cases}
		,
	\end{split}
\end{equation}
where $E_{1}\left(f_{ul}, l\right)$ and $E_{2}\left(f_{dl}, l\right)$ represent the channel estimation error, which can be modeled as additive white Gaussian noise (AWGN) with mean 0 and variance $\sigma_{E}^{2}$. After channel estimation, Alice and Bob get estimated CFRs $\mathbf{\hat{H}}_A=\{\hat{H}_{A}(f_{ul},0),...,\hat{H}_{A}(f_{ul},L-1)\}$ and  $\mathbf{\hat{H}}_B=\{\hat{H}_{B}(f_{dl},0),...,\hat{H}_{B}(f_{dl},L-1)\}$, respectively. \re{Different from \cite{yuan2020transfer} that used the channel impulse response (CIR) as channel information, we use the CFR to improve the key generation rate. Thus we could design a new neural network to fit our key generation problem.  }


\subsection{Feature Extraction}
Alice and Bob perform feature extraction to extract real-valued channel features $\mathbf{x}_A$ and $\mathbf{x}_B$ that are suitable for training the deep learning model and key generation. We can extract the magnitude and phase of CFR or directly separate the real and imaginary parts. In this paper, we extract the magnitude $\mathbf{x}'$ from $\mathbf{H}$ as the channel feature. \re{This is because multipath has a great impact on the phase of CFR, and it is difficult for the neural network to learn the mapping relationship between the phases of the uplink and downlink channels.}

The dataset is then normalized so that the range of the samples is between 0 and 1. The minimum and maximum values of the vectors in each dimension of the training dataset are saved and used for min-max normalization, i.e.,
\begin{equation}
\begin{split}
\mathbf{x}=\frac{\mathbf{x}'-\min (\mathbf{x}'_{\text{train}})}{\max (\mathbf{x}'_{\text {train}})-\min (\mathbf{x}'_{\text {train}})}, \quad \mathbf{x} \in[0,1]^{n^d},
\end{split}
\end{equation}
where $\mathbf{x}$ is the normalized value of $n^d$ dimensions.
\re{After feature extraction, Alice and Bob get suitable channel features $\mathbf{x}_A$ and $\mathbf{x}_B$, respectively.}

\subsection{Feature Mapping (only Alice)}
Based on \cite{zhang2022deep}, there is a feature mapping function $\mathcal{F}$ in each given environment. Alice can use $\mathcal{F}$ to predict the estimated downlink features $\mathbf{x}_{A}^{B}$ from $\mathbf{x}_A$, which can be expressed as
\begin{equation}
\mathbf{x}_{A}^{B} = \mathcal{F} (\boldsymbol{\Omega}, \mathbf{x}_A),
\label{mapping6}
\end{equation}
where $\boldsymbol{\Omega}$ is the parameters for feature mapping, which can be obtained by deep learning techniques. Through this step, Alice and Bob are considered to have obtained highly similar features $\mathbf{x}_{A}^{B}$ and $\mathbf{x}_B$, respectively. How to get the optimal value of parameters to minimize the gap between $\mathbf{x}_{A}^{B}$ and $\mathbf{x}_{B}$ is essential to generating highly similar features.

\subsection{Key Establishment}
Alice and Bob use obtained features to generate keys, including quantization, information reconciliation and privacy amplification \cite{li2019physical}. We use a Gaussian distribution-based quantization method with guard-band proposed in \cite{zhang2022deep} to get the initial keys $\mathbf{Q}_A$ and $\mathbf{Q}_B$. Denote the probability of the channel features $\mathbf{x}$ as a definite Gaussian distribution $\mathcal{N}_Q=\mathcal{N}(\mu,\sigma^2)$, where $\mu$ is the mean of vector $\mathbf{x}$, $\sigma$ is the standard deviation of vector $\mathbf{x}$, and $F^{-1}$ as the inverse of the cumulative distribution function (CDF) of $\mathcal{N}_Q$. The values between 0 and $F^{-1}(0.5-\varepsilon)$  are quantized as 0, and the values between $F^{-1}(0.5+\varepsilon)$ and 1 are quantized as 1.
The $\varepsilon \in (0,0.5)$ is defined as the quantization factor, and the values between $F^{-1}(0.5-\varepsilon)$ and $F^{-1}(0.5+\varepsilon)$ are discarded. In order to further improve the randomness of generated keys, Alice and Bob share the same random number seed to generate a random vector, and perform a random permutation on $\mathbf{Q}_A$ and $\mathbf{Q}_B$.

Information reconciliation and privacy amplification methods are then adopted to make Alice and Bob agree on the same key and remove any potential information leakage~\cite{li2019physical}. 
\re{In general, we can correct 25\% wrong bits of the initial keys. When the channel reciprocity is too poor and two parties cannot get similar channel features, the initial keys after quantization may have more than 25\% wrong bits. Even if the wrong bits are less than 25\% of the total keys, the larger wrong bits will cause a higher cost in information reconciliation. 
In this paper, we focus on improving channel reciprocity in FDD-OFDM systems, because it is essential to the success of key generation.}

\section{Problem Formulation and Algorithm Overview}
\label{system overview}

Among the four steps introduced in Section~\ref{Preliminary}, feature mapping is the crucial step for key generation between Alice and Bob. Therefore, this paper focuses on how to use deep learning techniques to quickly and efficiently obtain feature mapping functions $\mathcal{F}$ in multiple environments.


\subsection{Problem Statement}\label{Problem Formulation}
The performance of generated keys depends on the deep learning model. The existing works have verified the good fitting and generalization performance of the deep learning model to obtain the feature mapping function $\mathcal{F}$ in a certain environment \cite{zhang2021secret,wan2021secret,hou2021secret,zhang2022deep}. 
However, when the environment changes, the parameters of $\mathcal{F}$ are also affected by the environment, and the training samples and the actual samples of the deep learning model no longer obey a uniform distribution, which will lead to poor performance of the parameters in the new environment and even invalidate the effect of feature mapping. 

\begin{figure}[!t]
	\centering
	\includegraphics[width=0.8\linewidth]{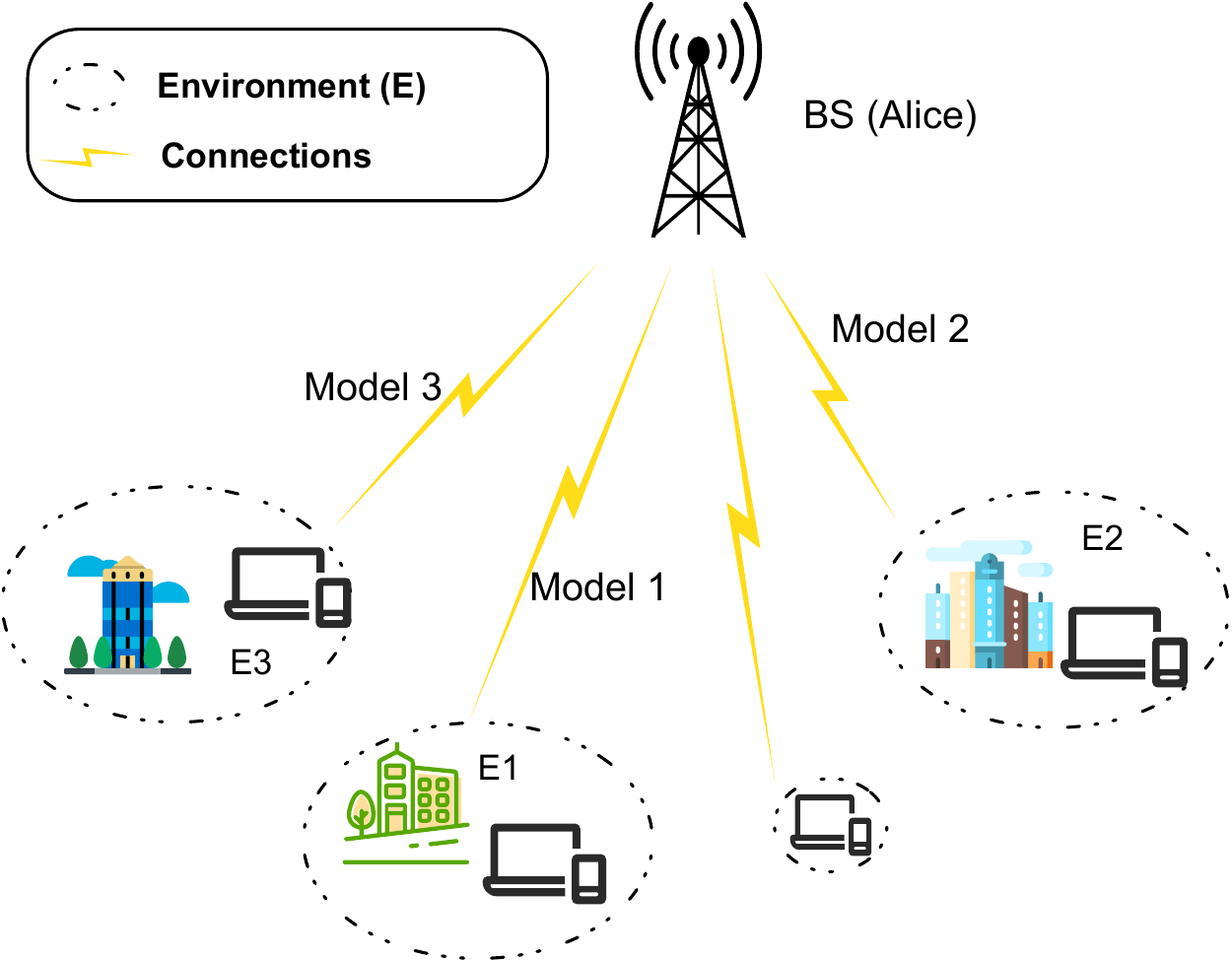}
	\caption{Deep learning-based PKG in multi-environments. }
	\label{DTL_se}
\end{figure}

As shown in Fig. \ref{DTL_se}, suppose a user is in the environment(E)1, a deep learning model 1 can be trained to get the parameters $\boldsymbol{\Omega}$ for feature mapping and key generation between the BS and the user in this given environment. However, when the user moves to other environments, such as E2 and E3, the training samples of model 1 and actual samples in the new environment no longer obey the same distribution, resulting in the performance of the pre-trained deep learning model being degraded or even invalid. 

A simple way to solve this problem is to re-collect the data and re-train the model for each new environment. However, training a model requires a lot of training data and training resources, which is unacceptable for practical applications. Therefore, this paper aims to address this problem and formulate it as a learning-based problem to more efficiently learn feature mapping functions in new environments using known knowledge in the source environment.


\subsection{Learning-based Problem for PKG}\label{DTL_problem}
Assume that there are data in $E$ wireless scenarios, and the uplink and downlink channel characteristics in the $e^{th}$ environment are defined as $\mathbf{x}_A^e$ and $\mathbf{x}_B^e$ respectively. According to \cite{nguyen2021transfer}, the "domain" and "task" in the $e^{th}$ environment are defined as following:

\begin{definition}[Domain]
The domain $\mathcal{D}(e)$ is composed of the feature space $\mathcal{X}^e$ and the marginal probability distribution $P({\mathbf{x}_A^e})$, i.e., $\mathcal{D}(e)=\{\mathcal{X}^e,P({\mathbf{x}_A^e})\}$. And the symbol $\mathcal{X}^e$ denotes an instance set, which is defined as all possible uplink
channel features $\mathbf{x}_A^e$, i.e., $\mathbf{x}_A^e \in \mathcal{X}^e$.
\end{definition}

\begin{definition}[Task]
The task $\mathcal{T}(e)$ is composed of the label space $\mathcal{Y}^e$ and a decision function $f^e$, i.e., $\mathcal{T}(e)=\{\mathcal{Y}^e,f^e\}$. And the symbol $\mathcal{Y}^e$ denotes an instance set, which is defined as all possible uplink
channel features $\mathbf{x}_B^e$, i.e., $\mathbf{x}_B^e \in \mathcal{Y}^e$. In other words, the task $\mathcal{T}(e)$ is the feature mapping from uplink to downlink.
\end{definition}

In a certain environment (domain $\mathcal{D}_{SE}$ and task $\mathcal{T}_{SE}$), the decision function $f^e$  can be obtained by model training. According to \cite{zhang2022deep}, the decision function $f^e$ can be considered as the feature mapping function $\mathcal{F}^e$ in the $e^{th}$ environment. Trained networks can act as the feature mapping function to achieve the feature mapping for key generation. However, when in a new environment (domain $\mathcal{D}_{TE}$ and task $\mathcal{T}_{TE}$), the feature mapping function $\mathcal{F}$ will change, and the performance of the trained model will be greatly reduced and cannot be used continuously.

We formulate this problem as a learning-based problem, i.e., learning from the known environments enables fast key generation in multiple new environments using a small amount of data and limited resources, formally defined as follows. Given the number of source tasks $E_S$, the source domains $\{\mathcal{D}_{SE}(e)\}_{e=1}^{E_S}$, the source tasks$\{\mathcal{T}_{SE}(e)\}_{e=1}^{E_S}$, the number of source tasks $E_T$, the target domains $\{\mathcal{D}_{TE}(e)\}_{e=1}^{E_T}$ and the target tasks $\{\mathcal{T}_{TE}(e)\}_{e=1}^{E_T}$, the learning-based problem in this paper is to leverage knowledge (data and models) from $\{\mathcal{D}_{SE}(e)\}_{e=1}^{E_S}$ and $\{\mathcal{T}_{SE}(e)\}_{e=1}^{E_S}$ to learn new tasks $\{\mathcal{T}_{TE}(e)\}_{e=1}^{E_T}$ with a small amount of data and limited resource, where $\{\mathcal{T}_{TE}(e)\}_{e=1}^{E_T}\ne\{\mathcal{T}_{SE}(e)\}_{e=1}^{E_S}$ and $\{\mathcal{D}_{TE}(e)\}_{e=1}^{E_T}\ne\{\mathcal{D}_{SE}(e)\}_{e=1}^{E_S}$. 

\subsection{Algorithm Overview}
\begin{figure}[!t]
	\centering
	\includegraphics[width=\linewidth]{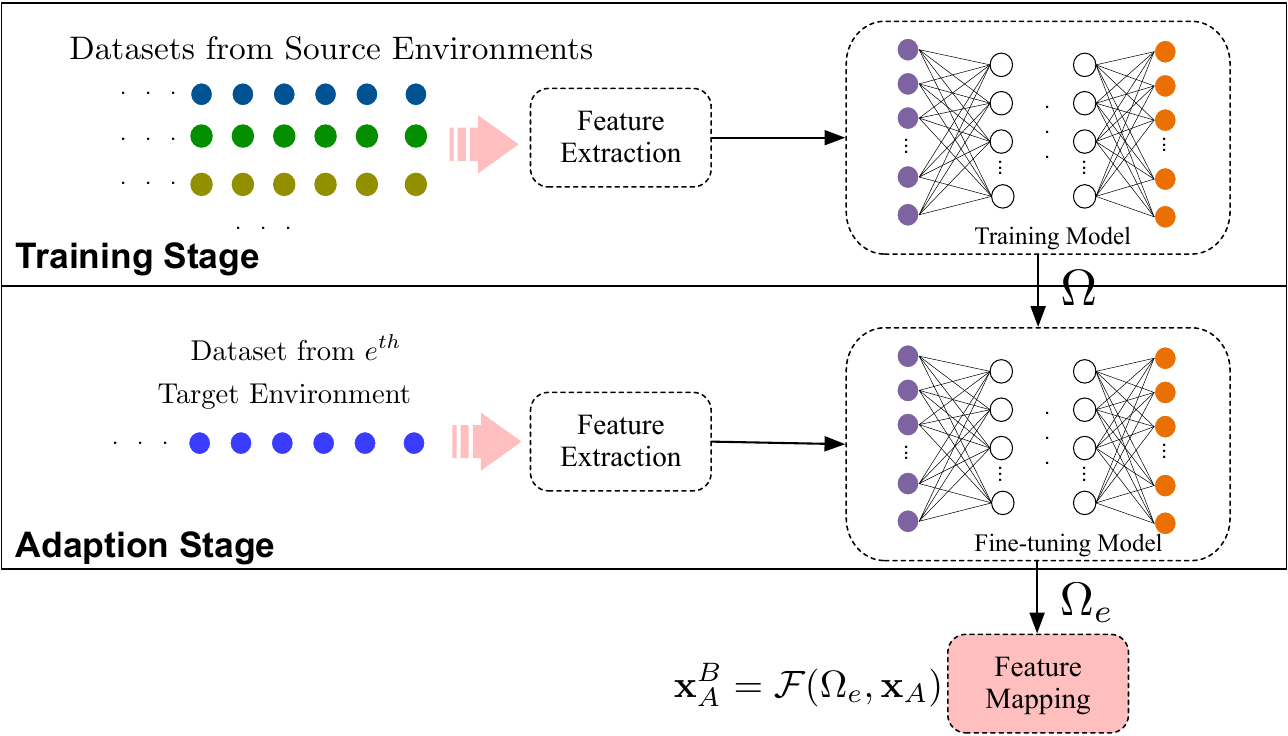}
	\caption{The proposed learning-based feature mapping scheme.}
	\label{learning_scheme}
\end{figure}

This paper proposes DTL-based and meta-learning-based feature mapping algorithms for key generation in multi-environments, elaborated in Section~\ref{DTL} and Section~\ref{meta}, respectively.
DTL and meta-learning aim to learn from source tasks to increase the generalization ability of the model under multi-task, and thus are two promising techniques for solving learning-based problems. 
Unlike learning functions $\mathcal{F}$ directly training the deep learning model in a given environment, these two algorithms include the training and adaptation stages, as shown in Fig.~\ref{learning_scheme}. 
\begin{itemize}
    \item Training stage: The two algorithms use datasets from known environments to train the model. DTL and meta-learning use different training methods, called pre-training and meta-training, respectively.
    \item  Adaptation stage: The two algorithms fine-tune the model using the datasets from the new environments, and then the fine-tuned model can be used for feature mapping and key generation.
\end{itemize}

This paper considers a simple FNN as the basic network structure to learn the feature mapping function $\mathcal{F}$ in the proposed algorithms, as shown in Fig. \ref{learning_scheme}. The input of the network is the uplink channel feature vector $\mathbf{x}_A$ obtained by Alice, and the output of the network is the result of the cascade of $\mathbf{x}_A$ through nonlinear transformation. The network is used to map the features of the uplink and downlink, so the output of the network is considered to be the estimated vector $\mathbf{x}_A^B$ of the downlink channel feature vector $\mathbf{x}_B$, which also can be expressed as \eqref{mapping6}, i.e.,  $\mathbf{x}_A^B=\mathcal{F}(\boldsymbol{\Omega},\mathbf{x}_{A})$,
where $\boldsymbol{\Omega}$ is all parameters in this network to be trained for feature mapping. The FNN consists of $M$ layers, including one input layer, $M-2$ hidden layers and one output layer. The output $f_m(\mathbf{x})$ of the $m^{th}$ layer is a nonlinear transformation of the output of $m-1^{th}$ layer, which can be written as:
\begin{equation}
\begin{split}
f_m(\mathbf{x})=F_{A,m}(\mathbf{W}_m \mathbf{x}+\mathbf{b}_m), 2 \leq m  \leq M,
\end{split}
\end{equation}
where $F_{A,m}$, $\mathbf{W}_m$ and $\mathbf{b}_m$ are the activation function of $m^{th}$ layer, weight vector between $(m-1) ^{th}$ and $m^{th}$ layers and bias vector of $m^{th}$ layer, respectively. The rectified linear unit (ReLU) function commonly used in regression problems is selected as the activation function $F_{A,m}$ of the hidden layers, and the sigmoid function is selected as the activation function $F_{A,m}$ of the output layer.

The purpose of the network is to learn the band feature mapping, so we could train a network to minimize the difference between network output $\mathbf{x}_A^B$ and $\mathbf{x}_{B}$. Because it is a vector regression problem, we consider using the mean squared error (MSE) as the loss function of the neural network. The loss function is defined as:
\begin{equation}
\begin{split}
\mathcal{L}_{\mathbb{D}}(\boldsymbol{\Omega})=\frac{1}{N_{batch}}\sum_{i=0}^{N_{batch}-1}\|{\mathbf{x}_{A}^{B}}(i)-\mathbf{x}_{B}(i)\|_2^2,
\label{loss}
\end{split}
\end{equation}
where $\mathbb{D}=\{(\mathbf{x}_{A},\mathbf{x}_{B})\}_{i=0}^{N_{batch}-1}$ is a batch-sized training dataset, $N_{batch}$ is the batch size.

\section{DTL-Based Feature Mapping}\label{DTL}
Based on the learning-based problem formulated in Section~\ref{DTL_problem}, this section proposes a DTL-based feature mapping to achieve key generation in new environments for FDD-OFDM systems. 
DTL transfers knowledge from the source environment to the target environment, so that the network in the target environment can achieve a better learning effect. In general, datasets in the source environments are abundant, while datasets in the target domains are small, so most DTL algorithms use datasets from source tasks to pre-train the model and then fine-tune it under a new task~\cite{nguyen2021transfer}. Like these works, in our proposed DTL-based feature mapping, we use the datasets from the source environments to pre-trained a model and then use a small number of samples to fine-tune the pre-trained model to obtain a model with good performance in the new environment.

\subsection{Definition of Dataset}
Assume that the source datasets $\{\mathbb{D}_{S}(e)\}_{e=1}^{E_S}$ is collected from $E_S$ source environments, where the dataset $\mathbb{D}_{S}(e)= \{(\mathbf{x}_A^{(n)}(e),\mathbf{x}_B^{(n)}(e))\}_{n=1}^{N_S}$ includes $N_{S}$ samples in the $e^{th}$ environment. 
Furthermore, it is necessary to collect datasets in multiple target environments to evaluate the performance of the algorithm. Assume that the target datasets $\{\mathbb{D}_{T}(e)\}_{e=1}^{E_T}$ from $E_T$ target environments, where the dataset in the $e^{th}$ environment $\mathbb{D}_{T}(e)=\{( \mathbf{x}_A^{(n)}(e),\mathbf{x}_B^{(n)}(e))\}_{n=1}^{N_T}$ includes $N_{T}$ data samples.


In the DTL algorithm, the datasets $\{\mathbb{D}_{S}(e)\}_{e=1}^{E_S}$ from all source environments are considered as a whole as the training dataset $\mathbb{D}_{Tr}$. The dataset $\mathbb{D}_{T}(e)$ in the target $e^{th}$ environment divides into adaption dataset $\mathbb{D}_{Ad}(e)=\{(\mathbf{H}_A^{(n)}(e),\mathbf{H}_B^{(n)}(e))\}_{n=1}^{N_{Ad}}$ and testing dataset $\mathbb{D}_{Te}(e)=\{(\mathbf{H}_A^{(n)}(e),\mathbf{H}_B^{(n)}(e))\}_{n=1}^{N_{Te}}$, where $N_{Ad}+N_{Te}=N_T$.

\subsection{Training (Pre-training) Stage}

The pre-training stage trains the model using dataset $\{\mathbb{D}_{S}(e)\}_{e=1}^{E_S}$ from the source environments to minimize the loss function $\mathcal{L}_{\mathbb{D}_{Tr}}(\boldsymbol{\Omega})$. 


In each batch, $N_{batch}$ samples are randomly selected from $\mathbb{D}_{Tr}$ to construct a batch training dataset and then ADAM \cite{kingma2014adam} optimizer is used to optimize the parameters of the model. When the performance of the model tends to be constant or the number of iterations reaches the upper limit, the parameters $\boldsymbol{\Omega}$ of the pre-trained model are obtained.

\subsection{Adaption Stage}
\label{adaption}
For the $e^{th}$ target environment, the parameters $\boldsymbol{\Omega}$ of the pre-trained model are used to initialize the network model parameter $\boldsymbol{\Omega}_e$ in the target environment. Then the parameter $\boldsymbol{\Omega}_e$ is optimized using the adaption dataset $\mathbb{D}_{Ad}(e)$ in the target environment to minimize $\mathcal{L}_{\mathbb{D}_{Ad}}(\boldsymbol{\boldsymbol{\Omega}})$. When the performance of the model tends to be constant or the number of iterations reaches the upper limit, the parameters $\boldsymbol{\Omega}_e$ of the model in a new environment are obtained.


After repeating the adaption stage in $E_T$ target environments, we can obtain the parameter $\{\boldsymbol{\Omega}_e\}_{e=1}^{E_T}$ in the target environments.
After this, the network parameter $\boldsymbol{\Omega}_e$ is fixed, and the network can be directly used in the feature mapping step in the target environment. Two users, Alice and Bob, follow the steps in Section~\ref{Preliminary} for key generation, where Alice uses the deep learning model with parameter $\boldsymbol{\Omega}_e$  for feature mapping.

We also calculate the average values of Normalized Mean Square Error (NMSE), KER, and Key Generation Rate (KGR) using the testing dataset $\{\mathbb{D}_{Te}(e)\}_{e=1}^{E_T}$ in target environments to evaluate the performance of the proposed algorithm.


\section{Meta-Learning-Based Feature Mapping}
\label{meta}
To better leverage knowledge from the source environments, this section proposes a meta-learning-based feature mapping. Most existing meta-learning algorithms are problem-specific. In order to eliminate the limitation of the model architecture on the application of meta-learning, a model-agnostic meta-learning (MAML) algorithm was proposed in \cite{finn2017model}. The goal of the algorithm is to achieve adaptation by alternately learning the parameter initialization of the model between the intra-task process and the cross-task process \cite{thrun1998learning}. 
Different from the DTL algorithm, the meta-learning algorithm requires training the model from multiple source tasks and aims to learn the best model initialization parameters through intra-task and cross-task updates. More importantly, unlike the DTL algorithm that emphasizes performance on current tasks, the meta-learning algorithm focuses more on the performance of new tasks.

\subsection{Definition of Dataset}

In meta-learning, the training dataset $\mathbb{D}_{Tr}$ is the combination of all datasets from the source environments $\{\mathbb{D}_{S}(e)\}_{e=1}^{ E_S}$, and the training dataset in each task is the dataset in each source environment. The $e^{th}$ task of training dataset $\mathbb{D}_{S}(e)$ needs to be divided into support dataset $\mathbb{D}_{Su}(e)$ and query dataset $\mathbb{D}_{Qu}(e)$, and $\mathbb{D}_ {Su}(e) \cap \mathbb{D}_{Qu}(e) = \emptyset$.
The dataset $\mathbb{D}_{T}(e)$ in the target environment is to be divided into adaptation dataset $\mathbb{D}_{Ad}(e)=\{(\mathbf{x}_A^ {(n)}(e),\mathbf{x}_B^{(n)}(e))\}_{n=1}^{N_{Ad}}$ and testing dataset $\mathbb{D} _{Te}(e)=\{(\mathbf{x}_A^{(n)}(e),\mathbf{x}_B^{(n)}(e))\}_{n=1 }^{N_{Te}}$, where $N_{Ad}+N_{Te}=N_T$.

\subsection{Training (Meta-training) Stage}
During the meta-training phase, the goal of the meta-learning algorithm is to learn a network initialization that can effectively adapt to new tasks. The underlying network architecture used here is the same as used in DTL. First, the parameters $\boldsymbol{\Omega}$ are randomly initialized and then updated through two iterative processes, namely intra-task update and cross-task update. The network parameters of each source task are optimized within the intra-task update, and the global neural network is optimized within the cross-task update.

\subsubsection{Intra-task Update}
A batch of $E_{batch}$ tasks is randomly selected from $E_S$ environments in a batch.
The goal of each task is to optimize its own neural network parameters on its support dataset $\mathbb{D}_{Su}(e)$. The objective of each task is achieved by minimizing the loss function based on supervised learning. $\boldsymbol{\Omega}_{S,e}$ is initialized as the global network parameter $\boldsymbol{\Omega}$. The objective function of each task can be expressed as:
\begin{equation}
\begin{split}
\boldsymbol{\Omega}_{S,e}=\arg \min _{\boldsymbol{\Omega}_{S,e}} \mathcal{L}_{\mathbb{D}_{Su}(e)}\left(\boldsymbol{\Omega}_{S,e}\right), \quad e=1, \ldots, E_{batch},
\end{split}
\label{}
\end{equation}
where $\boldsymbol{\Omega}_{S,e}$ is the network parameter of the $e^{th}$ task in the source task set. 
In each task, $\boldsymbol{\Omega}_{S,e}$ is initialized to $\boldsymbol{\Omega}$, and is then updated with $G_{Tr}$ times of gradient descent, i.e.,
\begin{equation}
\begin{split}
\boldsymbol{\Omega}_{S, e} \leftarrow \boldsymbol{\Omega}_{S, e}-\alpha \nabla\mathcal{L}_{\mathbb{D}_{\mathrm{Su}}(e)}\left(\boldsymbol{\Omega}_{S, e}\right),
\end{split}
\label{update}
\end{equation}
where $\alpha$ is the learning rate between tasks. The $\boldsymbol{\Omega}_{S, e}$ also can be updated by ADAM optimizer \cite{kingma2014adam}.

The intra-task update only performs once. In the original MAML algorithm \cite{finn2017model}, intra-task updates were made also only once, but some literature proposed to increase the times of intra-task updates to improve the performance~\cite{yuan2020transfer}. This paper analyzes the impact of task update times on performance in Section \ref{The Impact of Hyper-parameters in Meta-learning}. The results show that the increase of $G_{Tr}$ has no obvious effect on performance, but will increase the training cost, thus we set $G_{Tr}$ to 1.


\subsubsection{Cross-task Update}
The global network parameters $\boldsymbol{\Omega}$ are optimized based on the sum of the loss functions of all tasks in one batch. After the intra-task update, the loss function for all tasks in the batch can be estimated based on the related tasks and their query datasets $\{\mathbb{D}_{Qu}(e)\}_{e=1}^{E_{batch}}$. These loss functions can be added together to form the loss function used to optimize the global network parameters, i.e.
\begin{equation}
\begin{split}
 \mathcal{L}_{total}(\boldsymbol{\Omega})=\sum_{e=1}^{E_{batch}} \mathcal{L}_{\mathbb{D}_{Qu}(e)}\left(\boldsymbol{\Omega}_{S, e}\right).
\end{split}
\label{total}
\end{equation}
This loss function can also be minimized by optimizing $\boldsymbol{\Omega}$ by gradient descent or ADAM algorithm (learning rate $\beta$).

After the cross-task update is over, assign the updated $\boldsymbol{\Omega}$ to $\boldsymbol{\Omega}_{S,e}$, and then repeat the intra-task update and cross-task update until $ \mathcal{L}_{total}(\boldsymbol{\Omega})$ does not converge.  At this time, the parameter initialization of network learning is obtained, so that only a small number of samples can be adapted to the new environment.

It is clear that the training methods of the DTL algorithm and the meta-learning algorithm are almost completely different. In the DTL algorithm, the DTL algorithm minimizes the loss of the current model (only one) on all tasks, so the DTL algorithm hopes to find an initialization parameter that performs better on all current tasks. 
The meta-learning algorithm first uses the support dataset to minimize the loss function in each task, then uses the query dataset to minimize the loss sum of all tasks, and finally updates all model parameters with the model parameters obtained by minimizing the sum of loss functions of all tasks, which means that the performance of the model obtained after training to convergence under each task using the final initialization parameters obtained by meta-learning should still be as good as possible. Therefore, compared to the DTL algorithm, the meta-learning algorithm makes better use of knowledge in multiple environments, and the resulting model initialization parameters have better generalization, which is also proved in the results in Section \ref{simulation}.

\subsection{Adaption Stage}
This step is the same as the Section \ref{adaption}. We also use the fixed parameters $\{\boldsymbol{\Omega}_e\}_{e=1}^{E_T}$ for feature mapping and key generation to calculate evaluation metrics that can evaluate the performance of the proposed algorithm using the testing datasets $\{\mathbb{D}_{Te}(e)\}_{e=1}^{E_T}$ in the target environments.




\section{Theoretical Analysis of DTL and Meta-Learning}
\label{theoretical}
In this section, we first analyze channel differences in different environments, and then theoretically analyze which algorithm, i.e., DTL or meta-learning, may have better performance on this basis.

\subsection{Channel Difference Between Different Environments}
According to (\ref{CFR}), CFR is mainly affected by two factors, frequency and propagation environment. The previous work analyzed the influence of frequency in the same propagation environment \cite{zhang2022deep}. This paper mainly analyzes the channel gap in different propagation environments. The propagation environment refers to the physical environment in which the signal is transmitted, including communication distance, transmission medium, obstacles and other factors. In different propagation environments, during the transmission process, the signal reaches the receiving end through multiple paths with different transmission media, resulting in great changes in the phase and amplitude of the channel. In addition, since the phase of the channel may undergo various complex changes during signal transmission, such as phase mutations caused by multipath effects, the obtained channel amplitude will be more accurate in comparison. Therefore, this paper uses the channel amplitude as the channel feature for key generation.

We compare the data collected in the indoor and outdoor environments in reality, and the data collection process will be explained in Section~\ref{practical}. Kolmogorov-Smirnov test (K-S test) \cite{massey1951kolmogorov} is used to test whether the same subcarrier in two environments has the same data distribution. Assume that the $N_m$ sets of data measured on the $l^{th}$ subcarrier in indoor and outdoor environments are $\{H_{indoor}^{l,n}\}_{n=1}^{N_m}$ and $\{H_{outdoor}^{l,n}\}_{n=1}^{N_m}$, the test statistic (KS value) $K$ can be calculated as 
\begin{equation}
\begin{split}
K = \max\limits_{x}|\mathsf{F}_1(x) - \mathsf{F}_2(x)|,
\end{split}
\label{}
\end{equation}
where $\mathsf{F}_1(x)$ and $\mathsf{F}_2(x)$ are the empirical distribution functions (ECDFs) of the $\{H_{indoor}^{l,n}\}_{n=1}^{N_m}$ and $\{H_{outdoor}^{l,n}\}_{n=1}^{N_m}$, respectively. The p-value can be obtained by looking up the K-S test table. We have 900 samples in each subcarrier. At the Level of significance of 0.05, when the p-value is less than 0.04527, it is considered that two independent samples do not come from the same distribution. 
By performing the K-S test on all subcarriers in different environments, the p-values of all subcarriers are much less than 0.04527. Therefore, the data distribution on all subcarriers in different environments does not come from the same distribution. 

\begin{figure}[!t]
	\centering
	\includegraphics[width=\linewidth]{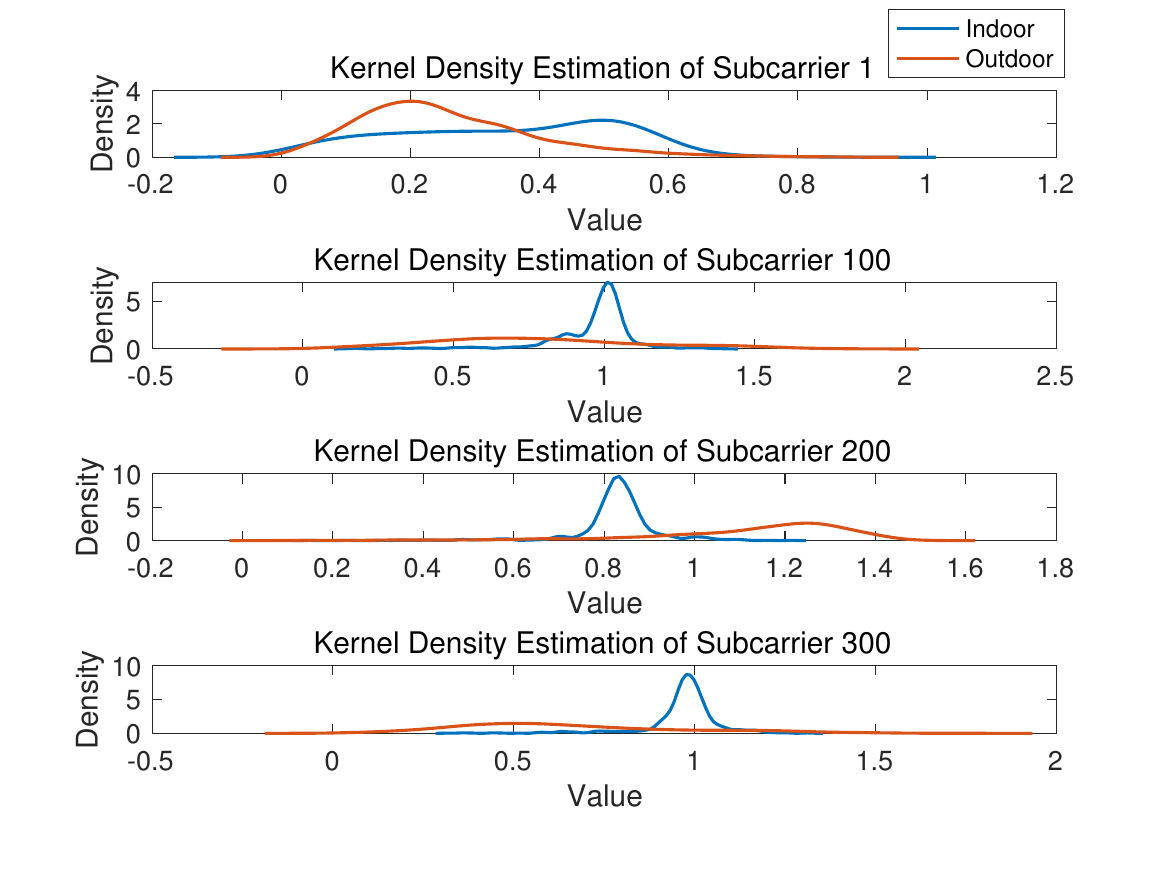}
	\caption{Comparison of kernel density estimation in different environments. } 
	\label{kernal}
\end{figure}
We further compare the kernel density estimation (KDE)~\cite{terrell1992variable} of values on the same subcarriers in different environments. 
KDE uses a kernel function (usually a Gaussian function) to smooth the empirical distribution of a set of data points to estimate the underlying probability density function. The output of the density function can be calculated by
\begin{equation}
\begin{split}
F_D(x) = \frac{1}{N_m b_w}\sum_{i=1}^{N_m} \mathcal{K}\left(\frac{x-x_i}{h}\right), 
\end{split}
\label{}
\end{equation}
where $x_i$ is the $i^{th}$ data point, $\mathcal{K}(u) = \frac{1}{\sqrt{2\pi}}e^{-\frac{u^2}{2}}$ is the kernel function, $b_w = 1.06\sigma N_m^{-1/5}$ is the bandwidth parameter, and $\sigma$ is the standard deviation of the sample data.

As shown in Fig. \ref{kernal}, the value distribution of the channel under different environments on the same subcarrier is basically completely different. This result also reflects the huge influence of the environment on the channel, so the model trained in one environment is difficult to use directly in the new environment.

\subsection{DTL v.s. Meta-Learning}
In order to better understand the DTL and meta-learning, we first analyze the loss function. The loss function in the training stage of DTL at the $t^{th}$ time step is calculated as
\begin{equation}
\begin{split}
&\mathcal{L}_{DTL}^{(t)}=\mathcal{L}_{\mathbb{D}_{tr}}\left(\boldsymbol{\Omega}^{(t-1)}\right),
\end{split}
\label{}
\end{equation}
where ${\mathbb{D}_{tr}}= \mathbb{D}_ {Su} \cup \mathbb{D}_{Qu}$.
According to (\ref{update}) and (\ref{total}), the loss function in the cross-task update of meta-learning at the $t^{th}$ time step is calculated as
\begin{equation}
\begin{split}
&\mathcal{L}_{total}^{(t)}=\sum_{e=1}^{E_{batch}} \mathcal{L}_{\mathbb{D}_{Qu}(e)}\left(\boldsymbol{\Omega}_{S, e}^{(t)}\right) \\
= &\sum_{e=1}^{E_{batch}}\mathcal{L}_{\mathbb{D}_{Qu}(e)}(\boldsymbol{\Omega}^{(t-1)}-\alpha \nabla\mathcal{L}_{\mathbb{D}_{\mathrm{Su}}(e)}\left(\boldsymbol{\Omega}_{S, e}^{(t-1)}\right)).
\end{split}
\label{}
\end{equation}

Through comparing the loss functions between DTL and meta-learning, we find that meta-learning not only takes into account the need to minimize the performance of the current query set but also considers the performance of the support set, while DTL directly considers all training set performance. This allows meta-learning to have better generalization capabilities and can better adapt to the data in the new environment whose distribution is different from the source dataset \cite{richa2023sharing}. According to the analysis in the previous section, the data distribution in different environments is completely different, so meta-learning can achieve better performance. This result is also verified in the following sections.

\re{We would like to emphasize that although the meta-learning algorithm performs better than the DTL algorithm in our simulation and experiment, the DTL still has the advantage of easy implementation and extension. Compared with the meta-learning algorithm, the DTL algorithm can be fine-tuned on the model of the current systems directly without a new pre-training process, and thus can directly extend the model currently trained in a single environment to multiple environments. }

\section{Simulation Evaluation}
\label{simulation}
In this section, we will first present the data generation and simulation setup. Then, we give the benchmarks, metrics and compare the performance of all algorithms.

\subsection{Simulation Setup and Dataset Generation}

\begin{figure}[!t]
	\centering
	\includegraphics[width=\linewidth]{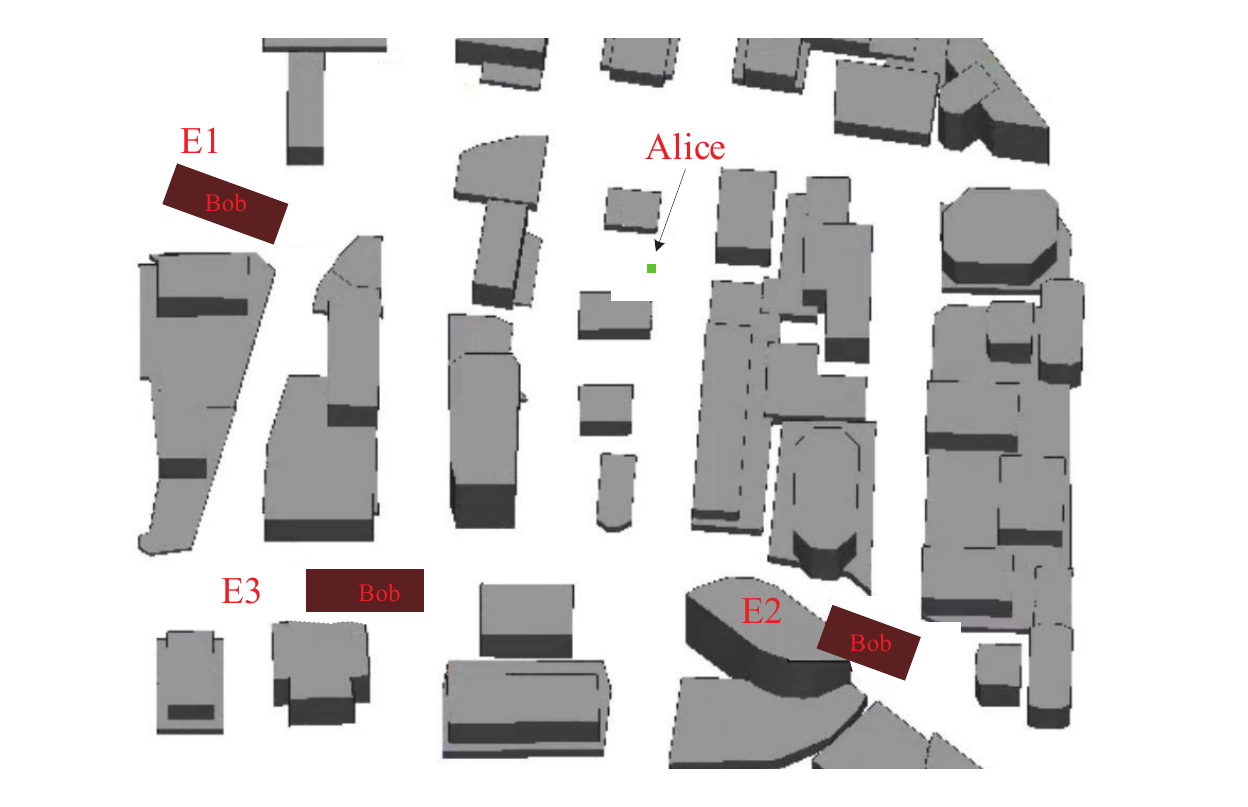}
	\caption{A overview of the ray-tracing outdoor scenario. }
	\label{scenario_urban}
\end{figure}

In the simulation, we consider multiple environments in the outdoor scenario, which is constructed based on the accurate 3D ray tracing simulator Wireless InSite~\cite{Remcom}.
An overview of the ray-tracing outdoor scenario is illustrated in Fig. \ref{scenario_urban}. The antenna of the base station (Alice) is located in a small green box with an outdoor height of 20 meters. The three maroon rectangles represent the possible positions of the user (Bob), and each rectangle represents an environment. 
The uplink channel and the downlink channel work on channels with frequencies of 2.4 GHz and 2.5 GHz, respectively. The number of OFDM subcarriers is 128 and the bandwidth is 20 MHz. 

Assuming that E1 is the source task environment and a total of 40,000 locations are collected. E2 and E3 are target task environments, and 5,000 locations are collected in each environment (80\% for adaptation and 20\% for testing). 
Since different environments are located in different areas, the scatterers and propagation paths are completely different, so the knowledge learned by the model is different, and the model trained in E1 may not be suitable for the new environment (E2 and E3). In response to this problem, this paper proposes two algorithms to use the collected data in E1 to obtain prior knowledge, so that the pre-trained model can quickly adapt to new environments (E2 and E3). 

A workstation with an Nvidia GeForce GTX 1660Ti GPU and an Intel Core I7-9700 CPU was used. This paper used Tensorflow 2.1 as the underlying framework of deep learning to build the network. The network parameters and some parameters in the training stage are shown in Table \ref{tab:4-setup}.
\begin{table}[!t]
\caption{Default Parameters of Proposed Algorithms} 
\label{tab:4-setup}
	\centering
	\begin{tabular}{|p{1.5cm}<{\centering}|p{3.6cm}<{\centering}| p{2.4cm}<{\centering}|}
		\hline
		& \bfseries Parameter & \bfseries Value  \\ \hline
	    \multirow{6}{*}{\shortstack{For All \\Algorithms}} &Number of neurons in hidden layers& (512,1024,1024,512)\\\cline{2-3}
	    &Batch size& 128\\\cline{2-3}
	    &Optimization & ADAM \cite{kingma2014adam} \\  \cline{2-3}
	    &Exponential decay rates for ADAM: ($\rho_1,\rho_2$) &(0.9,0.999)\\\cline{2-3}
	    &\re{Learning rate in Adaption Stage} &\re{1e-5} \\ \hline
 		\multirow{10}{*}{\shortstack{For\\ Meta-learning}} &Inner-task and across-task learning rate: $(\alpha, \beta)$& (1e-3,1e-3)\\\cline{2-3}
 		&The number of gradient update for inner-task training & 1\\\cline{2-3}
		&the number of the gradient update in fine-tuning and meta-adaption stages& 300\\\cline{2-3}
		&The number of source task in meta-learning& 400\\\cline{2-3}
 		&The number of samples in each source task& 100\\\hline
	\end{tabular}
\end{table}

\subsection{Benchmarks}
For comparison, we introduce two benchmarks, namely the direct algorithm and the joint dataset algorithm. All algorithms are explained below.
\begin{itemize}
\item[(1)] \re{\textit{Direct algorithm} represents the current deep learning-based key generation methods that ignore the feature mapping function changes in the multi-environments, which only include the training stage and testing stage. In the training phase, we use the training data from the source environment (E1) to minimize the loss function in (\ref{loss}) to optimize the deep learning model. In the testing phase, deep learning model performance is directly evaluated using data from new environments (E2 and E3) without adaptation. We test different model architectures in the recent works \cite{wan2021secret, zhang2022deep, chen2023physical}, and select the KGNet \cite{zhang2022deep} as our basic architecture.} 
\item[(2)]  \textit{Joint dataset algorithm} combines all the data in E1 and part of the data in E2 or E3 to form a joint training dataset and then uses the model trained by the joint training dataset to test the performance in E2 and E3, respectively \cite{yuan2020transfer}.
\item[(3)]  \textit{DTL algorithm} uses the proposed DTL-based feature mapping in Section~\ref{DTL} for key generation to test the performances.
\item[(4)]  \textit{Meta-learning algorithm} uses the proposed meta-learning-based feature mapping in Section~\ref{meta} for key generation to test the performances.
\end{itemize}

\re{For the direct algorithm, we select the optimal architecture for key generation to show that the deep learning model cannot be directly used in multi-environments. For the joint dataset algorithm, we directly combine the dataset from the source environments and the target environment, which can show that using suitable algorithms to adopt new environments is compulsory.}

For a fair comparison, some default training parameters adopted in all algorithms are consistent. Furthermore, the datasets used for training and adaptation in transfer learning and meta-learning algorithms are of the same size. The 40,000 total training dataset used in the DTL algorithm is divided into 400 datasets with a sample size of 100 in the meta-learning algorithm to represent the data under multiple tasks. In each task, the numbers of samples in the support dataset and query dataset are both 50.
The training dataset used by the joint dataset algorithm is the combination of the training dataset and the adaptation dataset in the transfer learning and meta-learning algorithms.

\subsection{Evaluation Metrics}
We use the following metrics for performance evaluation.
\begin{itemize}
	\item \textit{NMSE} is used to evaluate the predictive accuracy of the network, which  is defined as
\begin{equation}
\begin{split}
\mathrm{NMSE}=E\left[\frac{\parallel{\mathbf{x}}_{A}^{B}-\mathbf{x}_B\parallel_2^2}{\parallel \mathbf{x}_{B}\parallel_2^2}\right],
\end{split}
\label{NMSE}
\end{equation}
where $E\left[\cdot\right]$ represents the expectation operation. 
	\item \textit{KER} is defined as the number of error bits divided by the number of total key bits.
	\item \textit{KGR} is defined as the number of initial key bits divided by the number of subcarriers. 
	\item \textit{Randomness} reveals the distribution of bit streams. The National Institute of Standards and Technology (NIST) statistical test \cite{rukhin2001statistical} is used for the randomness test for the generated keys.
\end{itemize}

\subsection{The Impact of Hyper-parameters in Meta-learning}
\label{The Impact of Hyper-parameters in Meta-learning}
The selection of the number of iterations $G_{Tr}$ in the task and the batch size $E_{batch}$ in the training phase are very important to the meta-learning algorithm. These two parameters are analyzed below.

\begin{figure}[!t] 
    \centering 
    \includegraphics[width=\linewidth]{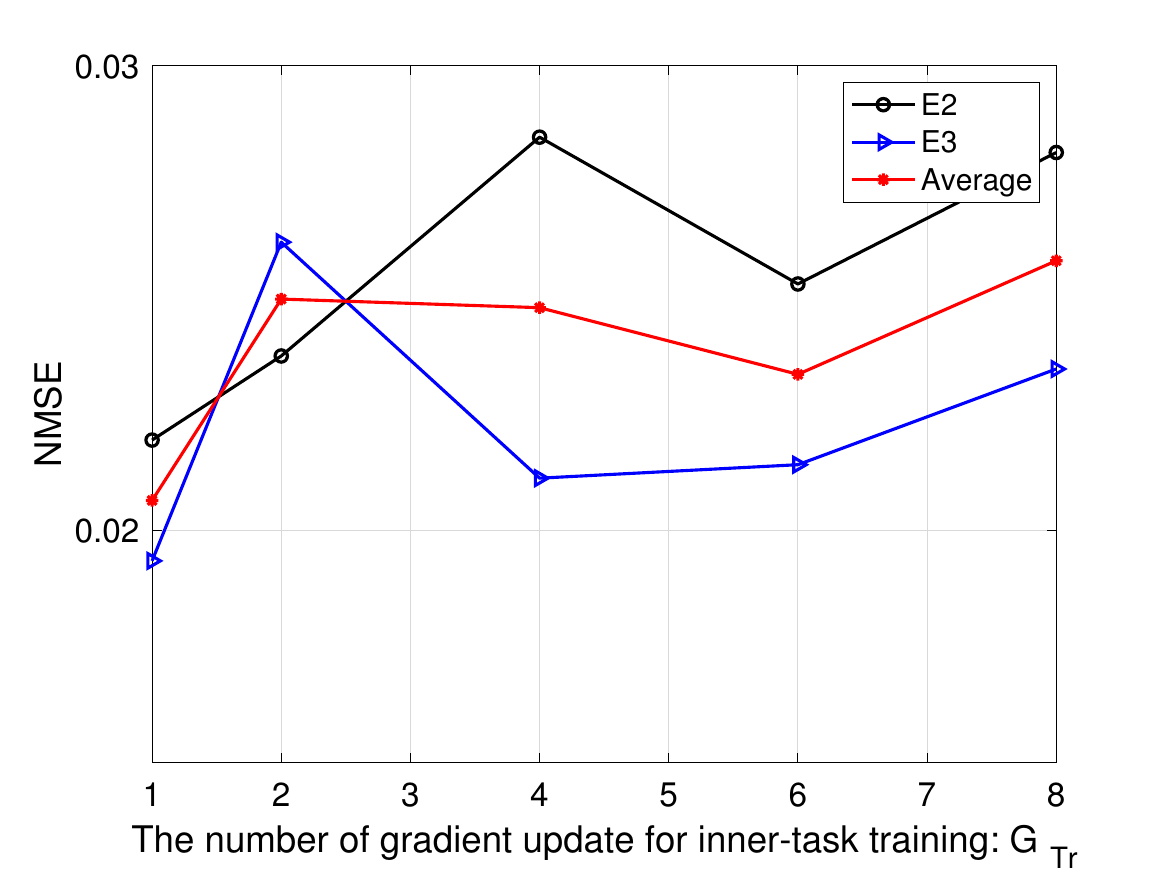} 
    \caption{The NMSE performance comparison for different numbers of iterations $G_{Tr}$.}
    \label{Gtr_NMSE} 
\end{figure}

For some tasks, the increase of $G_{Tr}$ can greatly improve the performance. For example, the work in~\cite{yuan2020transfer} sets $G_{Tr}$ to 3, which improves the downlink channel prediction accuracy in massive MIMO systems. At the same time, as $G_{Tr}$ increases, more memory and time resources are required for meta-learning training. Therefore, the value of $G_{Tr}$ should be determined comprehensively by weighing the consumed resources and performance. In this paper, $G_{Tr}$ is set as \{1, 2, 3, 4, 6, 8\} for learning, and tests are carried out in the outdoor environment respectively. The results are shown in Fig. \ref{Gtr_NMSE}. The results show that with the increase of $G_{Tr}$, the performance of the meta-learning algorithm does not improve, but basically stabilizes around a certain range. Therefore, in order to guarantee the minimum resource consumption, the $G_{Tr}$  is set to 1.

\begin{figure}[!t] 
    \centering 
    \includegraphics[width=\linewidth]{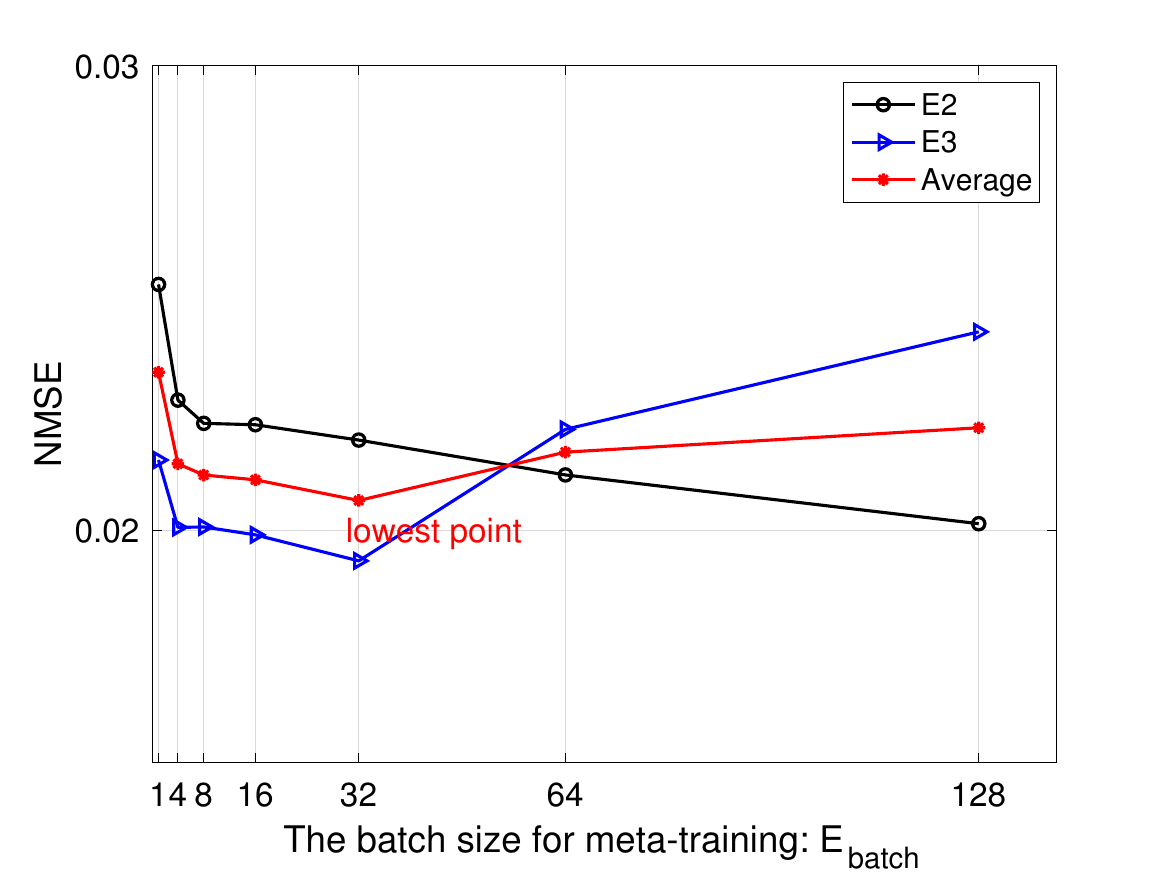} 
    \caption{The NMSE performance comparison for different numbers of the batch size $E_{batch}$.}
    \label{Kb_NMSE} 
\end{figure}

Reasonable selection of the batch size $E_{batch}$ in the training phase is also very important for the training effect. Since the choice of batch size $E_{batch}$ has nothing to do with the resource consumption of training, it is only necessary to focus on the training performance under different batch sizes. Fig.~\ref{Kb_NMSE} compares the NMSE performance under different batch sizes $E_{batch}$. The results show that the tested NMSE performance is getting better with the increase of $E_{batch}$ and reaches optimal when $E_{batch}=32$, which is used in the rest of the paper. 

\subsection{Performance of Reciprocal Features}

\begin{figure}[!t] 
	\centering 
	\includegraphics[width=\linewidth]{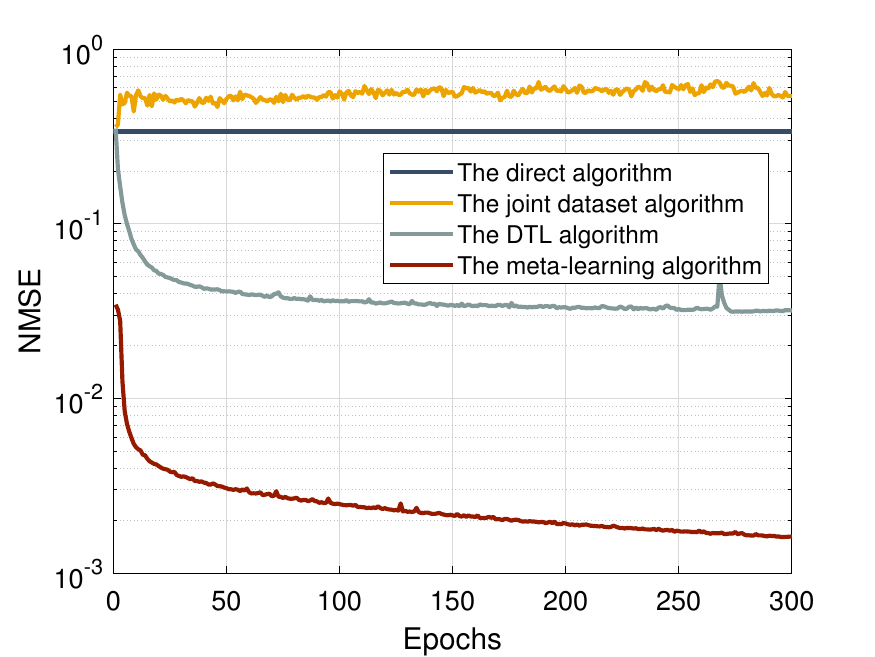} 
	\caption{\re{The NMSE performance during the adaption stage.}}
	\label{Gad_nmse}
\end{figure}

Fig. \ref{Gad_nmse} compares the NMSE performance of the four algorithms during the adaption stage in E2. Since the direct algorithm has no adaptation phase, it is set as a fixed value for its test results. The results show that the algorithms based on transfer learning and meta-learning are better than the direct and joint dataset algorithms. The NMSE of the joint dataset algorithm increases with the number of epochs. 
In the joint dataset, the number of data samples of E2 is much larger than that of E1, and overfitting occurred during the training process. In addition, the distribution of data samples in different environments is too different, so its test performance is weaker than the direct algorithm. This result suggests that it is necessary to skillfully utilize a small number of datasets in new environments, rather than simply superimposing the data directly.
In addition, the meta-learning algorithm is significantly better than the DTL algorithm.


\begin{figure}[!t] 
	\centering 
	\includegraphics[width=\linewidth]{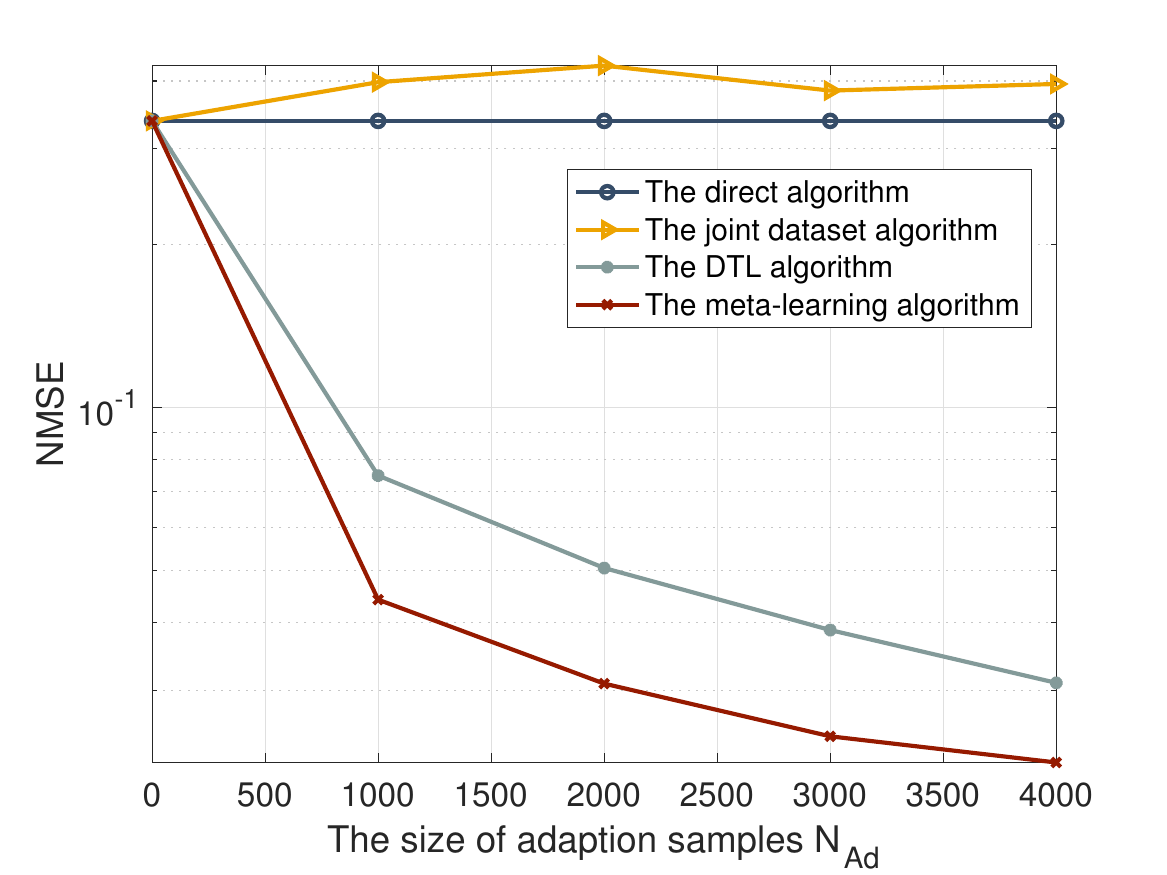} 
	\caption{The NMSE performance of the four algorithms versus the size of adaption samples $N_{Ad}$.}
	\label{Nad_nmse}
\end{figure}

Fig. \ref{Nad_nmse} compares the influence of the number of adaptation dataset samples $N_{Ad}$ on the performance of the four algorithms. Since the direct algorithm does not use the adaptation dataset in the new environments, it is assumed that the performance of the algorithm under different sample numbers is consistent. When the number of samples $N_{Ad}=4000$, since the data in the new environment in the joint dataset only accounts for $4000/41000$ of the total data, the resulting overfitting reaction makes the performance of the joint dataset algorithm is even worse than that of the direct algorithm. 
Overall, the meta-learning and DTL algorithms can achieve better performance than the two benchmarks with a smaller number of samples, and the performance of the meta-learning algorithm is better than that of the DTL algorithm.



\begin{figure}[!t] 
	\centering 
	\includegraphics[width=\linewidth]{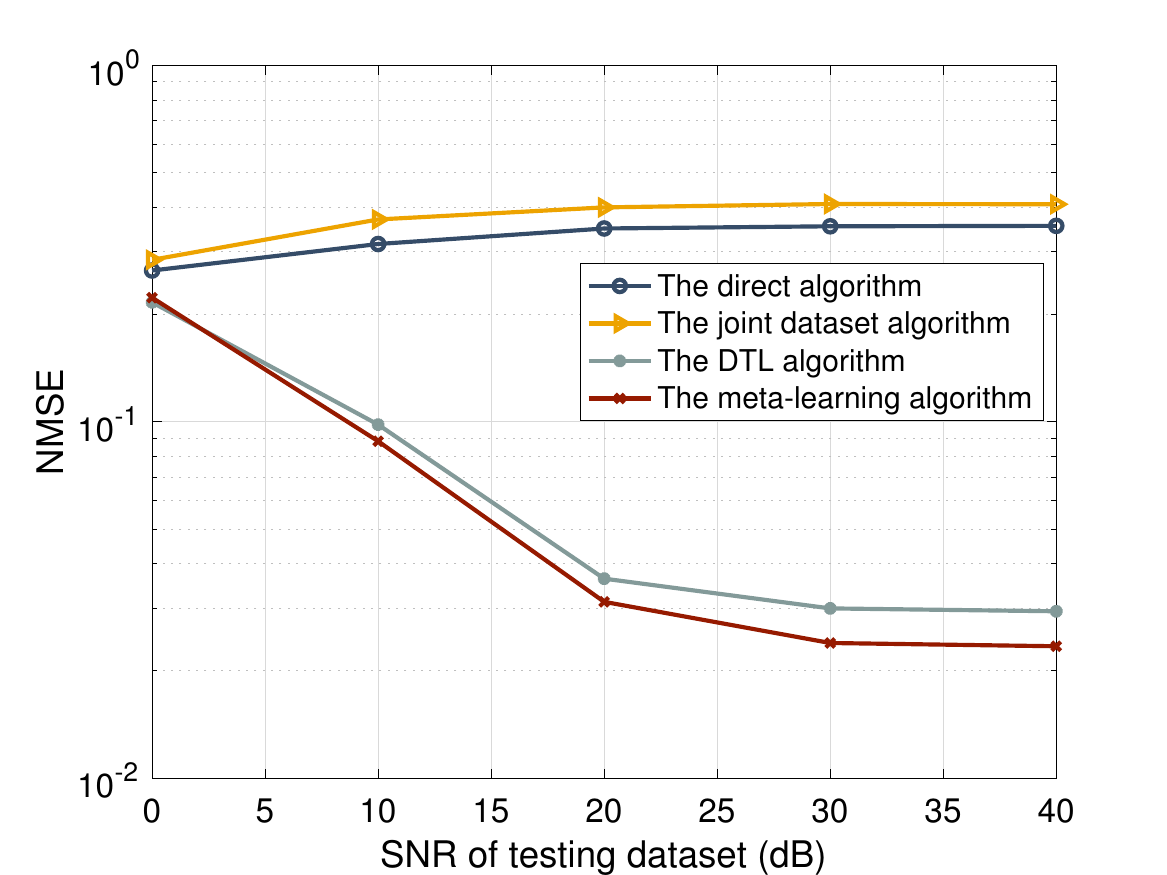} 
	\caption{The NMSE performance versus different SNRs.}
	\label{SNR_nmse_indoor}
\end{figure}

Testing datasets at SNRs of \{0, 10, 20, 30, 40\} dB are also generated to analyze the generalization performance of the four algorithms. Fig. \ref{SNR_nmse_indoor} compares the performance of the four algorithms tested under different SNRs. The results show that the DTL and meta-learning algorithms can achieve better performance than the direct and joint dataset algorithms. The NMSE is still less than 0.1 when the SNR is lower than 10 dB. By this time, the meta-learning algorithm can still effectively improve the reciprocity of features obtained by Alice and Bob.

\subsection{Performance of Initial Keys}

Based on the above analysis of the performance of the feature reciprocity generated by the algorithms, this section analyzes the performance of the initial keys, which includes KER, KGR, and key randomness. In the following section, the quantization factor $\varepsilon$ is set to 0.1, which means that 20\% of the features near the isolation zone are removed in the quantization.

\begin{figure}
	\centering
	\subfigure[The KER performance versus different testing SNRs.]{
		\begin{minipage}[b]{\linewidth}
			\includegraphics[width=\linewidth]{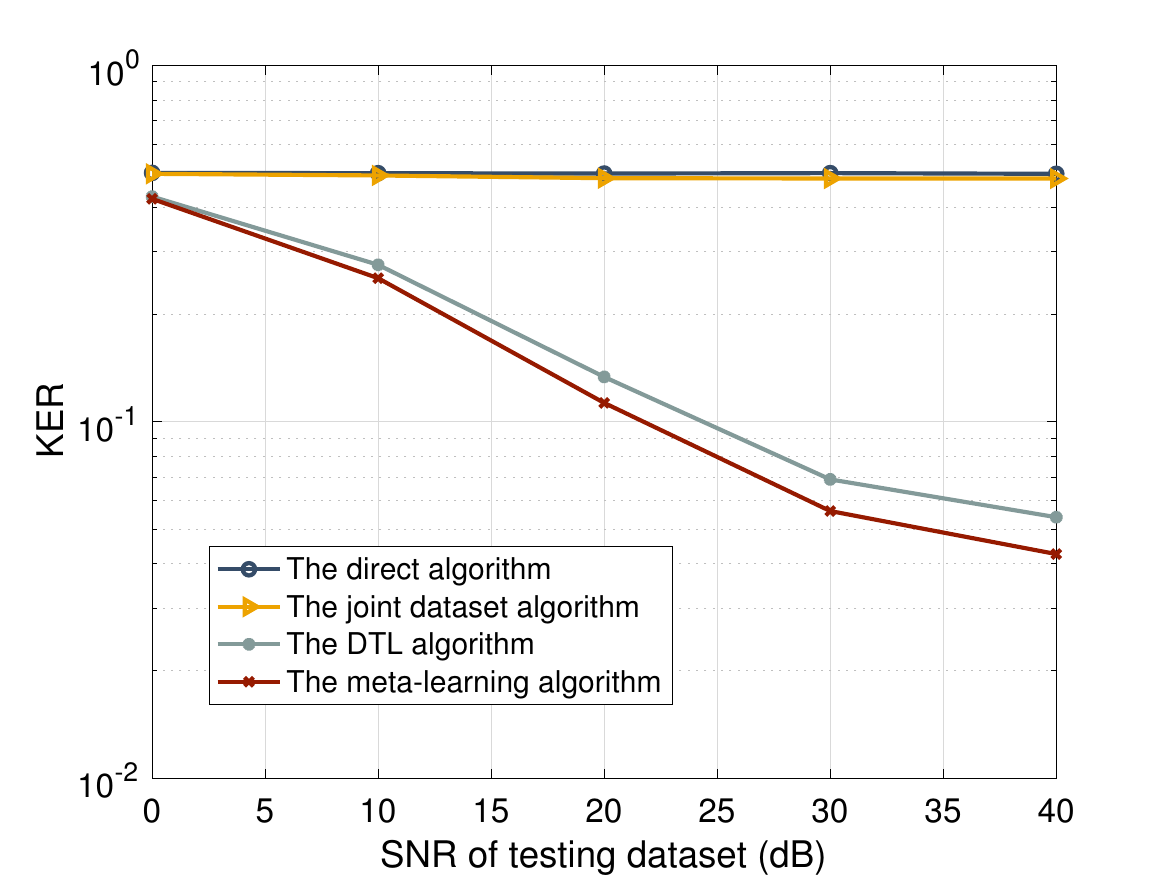}
		\end{minipage}
		\label{KER_indoor}
	}
	\subfigure[The KGR performance versus different testing SNRs.]{
		\begin{minipage}[b]{\linewidth}
			\includegraphics[width=\linewidth]{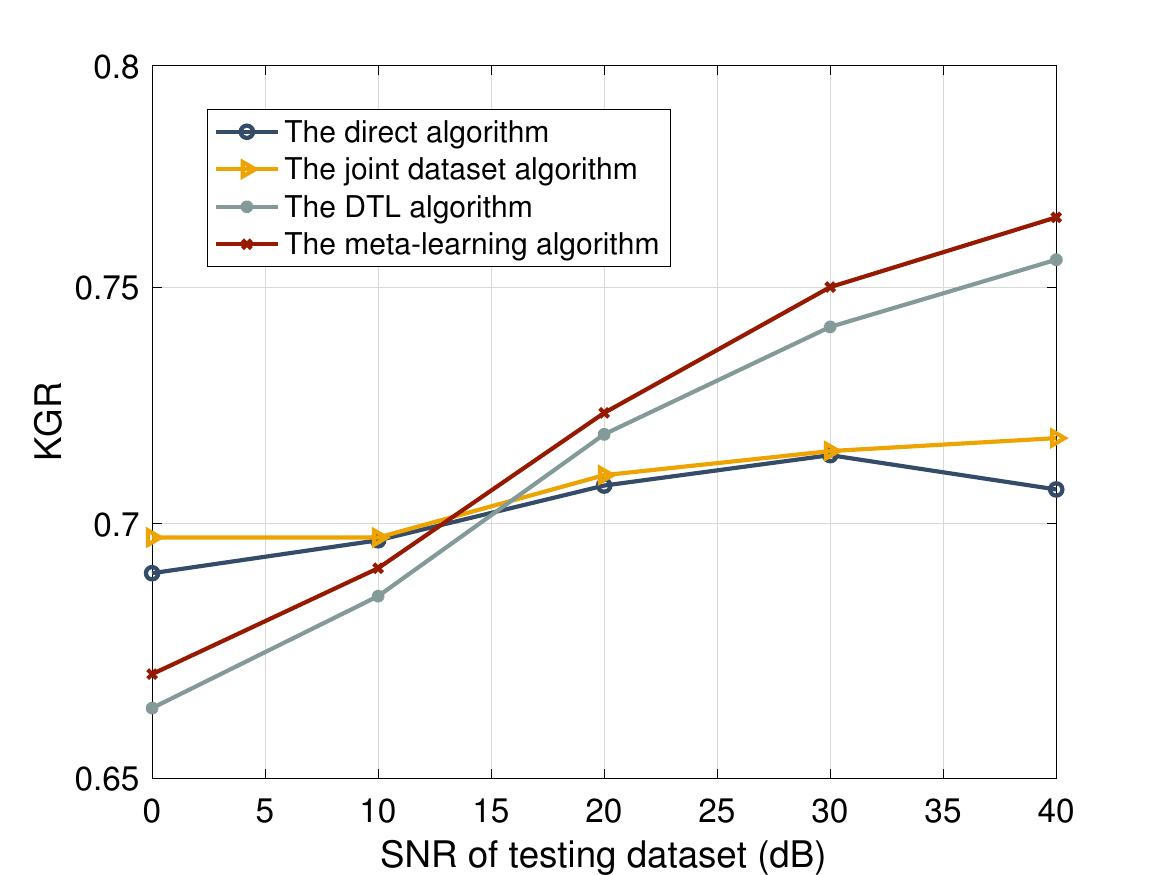}
		\end{minipage}
		\label{KGR_indoor}
	}
	\caption{The performance of initial keys under different SNRs of the testing dataset.}
	\label{}
\end{figure}

Fig. \ref{KER_indoor} and Fig. \ref{KGR_indoor} compare the performance of the keys generated by the four algorithms tested under different SNRs. As shown in Fig. \ref{KER_indoor}, the KERs of the keys generated by the direct and joint dataset algorithms are as high as 50\%. 
This indicates that the model trained in the source environments is invalid in the new environment. The DTL and meta-learning algorithms can significantly reduce the KERs of generating keys in these new environments. For example, the DTL algorithm and meta-learning algorithm generate keys at the SNR of 20 dB with the KER of 13.37\% and 11.3\%, respectively. 
As shown in Fig. \ref{KGR_indoor}, when the SNR is higher than 20 dB, the KGRs of the keys generated by DTL and the meta-learning algorithms are also higher than those of the keys generated by the other two benchmark algorithms. However, when the SNR is less than 15 dB, the obtained features of DTL and the meta-learning algorithms are more concentrated in the isolation zone, so the KGRs are lower than the other two benchmark algorithms.

%
%

The NIST test suite is used to test the randomness of the generated keys. Each test will return a p-value. When the p-value is greater than a commonly chosen threshold of 0.01, the generated key passes that test. A serial test is composed of two types of serial tests. When both tests pass, the serial test is considered to be passed. Table \ref{table_randomness_1} shows that all cases pass the test with p-values much larger than 0.01.
\begin{table}[!t] 
\caption{NIST statistical test results of the simulation data.} 
	\centering
	\begin{tabular}{|p{4cm}<{\centering}| p{2cm}<{\centering} |}
		\hline
		 \bfseries Test&\bfseries P-value \\ \hline
		 Approximate Entropy& 0.6606 \\\hline
		 Block Frequency & 0.0259 \\\hline
		 Cumulative Sums  &0.3752 \\\hline
		 Discrete Fourier Transform & 0.3019 \\\hline
		 Frequency & 0.4795 \\\hline
		 Ranking & 0.1371 \\\hline
		 Runs & 0.6897 \\\hline
		 \multirow{2}{*}{\shortstack{Serial }} & 0.7316 \\ &0.7237 \\
		 \hline
	\end{tabular}
\label{table_randomness_1}
\end{table}

\subsection{Complexity Analysis}

\begin{table}[!t] 
\caption{Complexity analysis of four algorithms.}
	\centering
	\begin{tabular}{|p{2.5cm}<{\centering}| p{1cm}<{\centering} | p{1cm}<{\centering} |p{1.5cm}|}
	\hline
		 \bfseries Algorithm &\bfseries Training Stage & \bfseries  Adaption Stage &\bfseries Key Generation Stage \\ 
		 \hline
		 Direct Algorithm & 183s & - & 0.95e-4s\\ 
		 \hline
		 Joint Training Algorithm & 253s  & - & 0.95e-4s\\ 
		 \hline
		 DTL Algorithm & 183s  &  37s & 0.95e-4s\\ 
		 \hline
		 Meta-learning Algorithm &\textbf{110s} &  \textbf{38s} & 0.95e-4s \\ 
		 \hline
	\end{tabular}
\label{table_complexity_1}
\end{table}
We compare the time cost of the four algorithms in Table~\ref{table_complexity_1}.
In the training stage, since the training process of meta-learning includes multiple intra-task and cross-task updates, meta-learning consumes significantly more resources than DTL. However, it was found experimentally that the training process of meta-learning requires only 10 iterations before the loss function stops decreasing, which takes about 110 seconds. 
The DTL and meta-learning algorithms increase the consumption required for the adaptation stage on top of the direct algorithm and the joint training algorithm, however, the improved performance shows that the consumption is worth it. In our experiments, the DTL and meta-learning algorithms only take about 37 seconds to complete the adaptation to the new environment, which is an acceptable cost.
In the key generation stage, the time cost of each feature mapping is around 0.95e-4 seconds, which can be done in almost real-time.

In summary, it takes about 148 seconds to train and adapt the network in total, and only 0.95e-4 seconds to use the network for feature mapping in the key generation stage. Furthermore, the training stage only needs to be performed once for all environments, and the adaptation stage is performed only once for each environment. Compared with the training consumption of networks used in other areas, the proposed algorithms can achieve fast key generation in FDD-OFDM systems. 

\re{As for the model size, the model used has 2,763,775 parameters and requires 20.5 MB of storage space, which is affordable to a BS. In addition, the inference stage is usually not computationally expensive, which can be handled by the BS too.}

\re{
\subsection{Security Analysis}
The security of our scheme can still be guaranteed even if the feature mapping function is leaked. When Eve is located over half of the wavelength away from Alice or Bob, she cannot get correlated channel measurements. In this circumstance, even though Eve has access to the feature mapping function, it cannot infer the channel features.}

\section{Experimental Evaluation}
\label{practical}

In this section, we will introduce a FDD-OFDM key generation hardware platform based on GNURadio and USRP, and then evaluate the performance of the proposed scheme in a real environment.

\subsection{Experimental Setup and Dataset Collection}
We built an experimental platform based on the GNURadio software radio suite and USRP N210 to collect FDD channel data in realistic scenarios to verify performance. Two USRP N210 SDR platforms \cite{ettus} are used as Alice and Bob to receive and transmit signals processed by MATLAB. OFDM probing signals are generated and processed in MATLAB and stored as data stream files in the PC.
In the experiment, it is difficult to guarantee that Alice and Bob transmit and estimate the probing signals at the same moment. First, Alice's transmitter sends a sounding signal to Bob, and Bob's transmitter is triggered to send an OFDM signal when Bob's receiver detects the sounding signal, and then Alice's receiver detects the sounding signal and notifies Bob to start subsequent channel estimation. This mechanism affects the accuracy of the CSI obtained by Alice and Bob to a certain extent, but due to the short communication time, it can be regarded as an FDD channel. In order to avoid the influence of the ISM band signal on the experiment, the uplink and downlink carrier frequencies $f_{ul}$ and $f_{dl}$ are set to 2.535 GHz and 2.435 GHz respectively, the transmit gain and receive gain are 30 dB, 
and the number of subcarriers is 511.

\begin{figure}
	
	\subfigure[Indoor.]{
			\begin{minipage}[b]{\linewidth}\centering
					\includegraphics[width=2.6in]{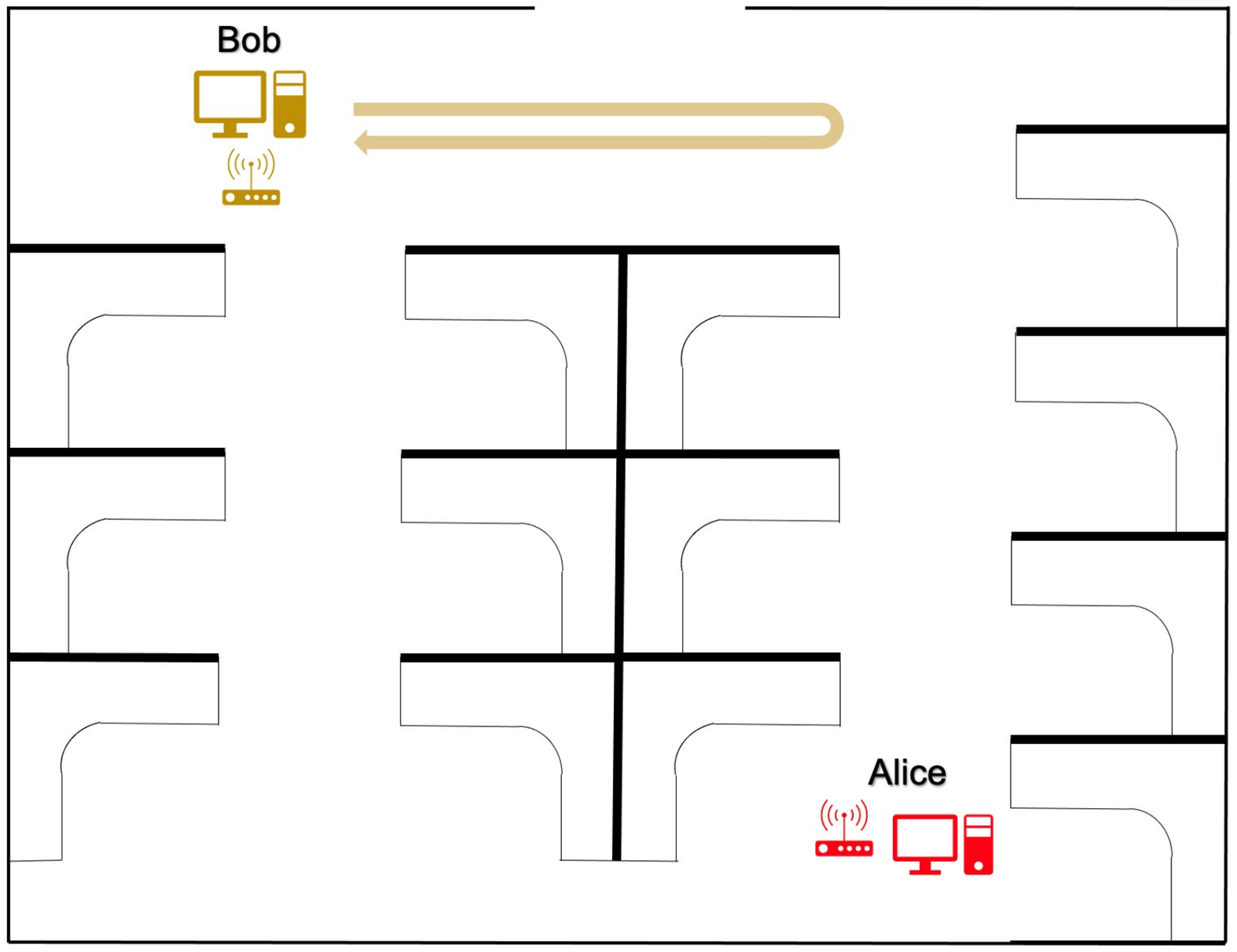}
				\end{minipage}
			\label{}
		}
		
	\subfigure[Outdoor.]{
			\begin{minipage}[b]{\linewidth}\centering
					\includegraphics[width=2.6in]{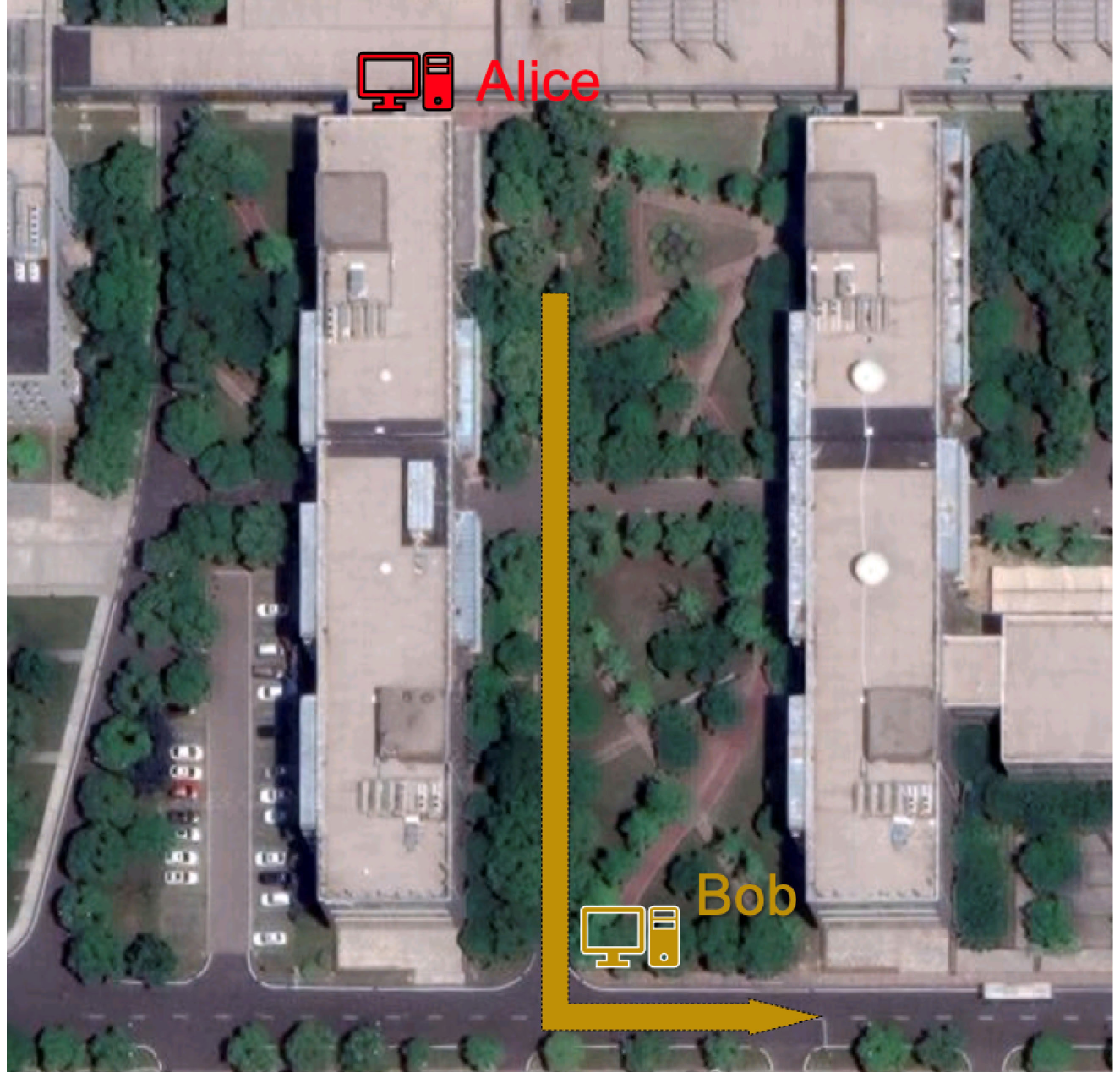}
				\end{minipage}
			\label{}
		}
	\caption{Floor plans of an indoor environment and an outdoor environment, where Alice remains stationary and Bob moves slowly along the brown-colored line. }
	\label{experient}
\end{figure}

We conducted extensive experiments in two scenarios, as shown in Fig. \ref{experient}, one is an interior scene of an office room, and the other is an outdoor scene of the Purple Mountain Laboratories, China. Alice remains stationary, Bob keeps moving slowly along the route marked in the diagram. There may also be other variations caused by people or vehicles moving around. 
We obtained more than 900 groups of uplink and downlink CSI vectors in both scenarios.

The amount of data collected in the experiment is less than that collected in the simulation, so we adjusted the batch size $N_{batch}$ and $E_{batch}$ in the transfer learning and meta-learning algorithms to 8, the number of samples in each task in meta-learning is 30. In each task, the numbers of samples in the support
dataset and query dataset are both 15. The number of layers, the number of neurons, and the optimizer of the deep learning model remain unchanged.

\subsection{Experimental Results}

First, we evaluate the performance of deep learning-powered FDD-OFDM key generation in a single environment, from which both training and test data are collected. We use 80\% of the data for training and 20\% of the data for testing in two scenarios. This is the first time the actual performance of such methods has been evaluated experimentally. 
The results are shown in Table~\ref{test_pratical}. In the indoor scenario, the KER before and after feature mapping is 0.3539, and 0.0681, respectively. In the outdoor scenario, the KER before and after feature mapping is 0.3683 and 0.0555 respectively. The experimental results show that the deep learning method can significantly improve the reciprocity in FDD systems and greatly improve the performance of the generated keys in a single environment. 
\begin{table}[!t]
\caption{Comparison of performance before and after feature mapping.}
\centering
\begin{tabular}{|p{1cm}<{\centering}|p{1cm}<{\centering}| p{2.4cm}<{\centering}|p{2.4cm}<{\centering}|}
\hline
\textbf{Scenario}& \textbf{Metric} & \textbf{Before Feature Mapping} & \textbf{After Feature Mapping} \\ \hline
\multirow{3}{*}{\shortstack{Indoor}} & NMSE    &       0.2031 &0.0840
\\ \cline{2-4} 
    & KER    &    0.3539     & \textbf{0.0681}\\ \cline{2-4} 
    & KGR    &    0.7583  &    0.8369     \\ \hline
\multirow{3}{*}{\shortstack{Outdoor}}  & NMSE    & 0.3158       & 0.0811
\\ \cline{2-4} 
& KER    &  0.3683   &   \textbf{0.0555} 
\\ \cline{2-4} 
& KGR    &  0.7144      &    0.7882    \\ \hline
\end{tabular}
\label{test_pratical}
\end{table}


\begin{figure}[htbp]\label{f_1}
\centering
\subfigure[The NMSE performance versus the number of adaption samples.] {
 \label{fig:cdf-1}     
\includegraphics[width=\columnwidth]{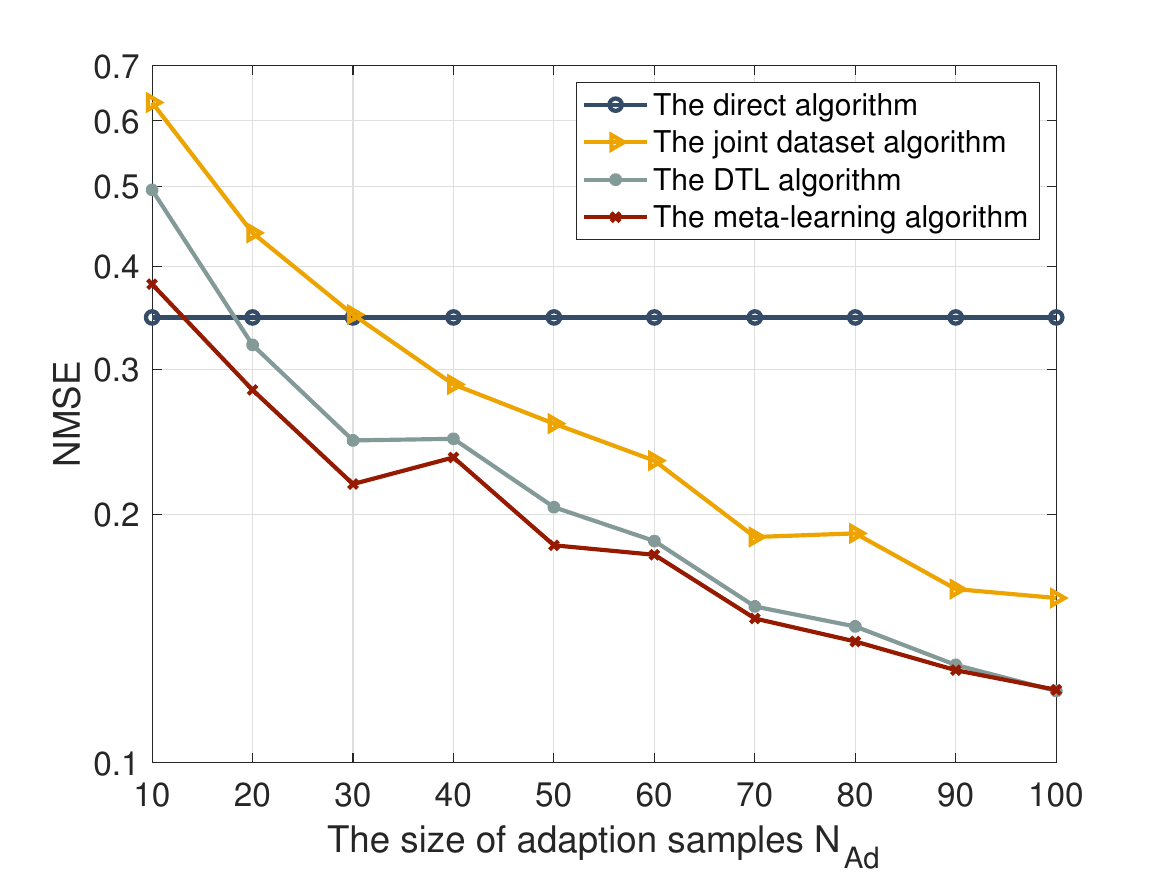}}

\subfigure[The KER performance versus the number of adaption samples.] {
 \label{fig:cdf-2}     
\includegraphics[width=\columnwidth]{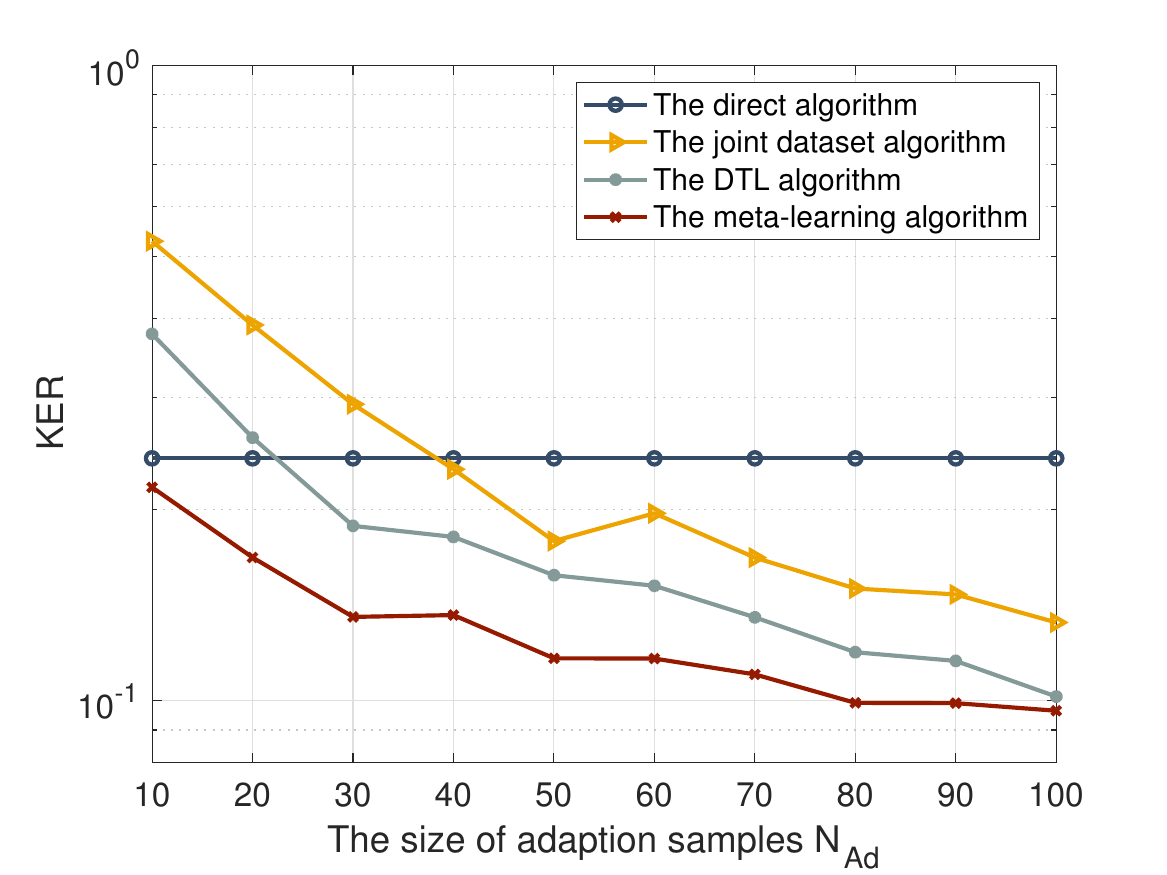}}  

\subfigure[The KGR performance versus the number of adaption samples.] {
 \label{fig:cdf-3}     
\includegraphics[width=\columnwidth]{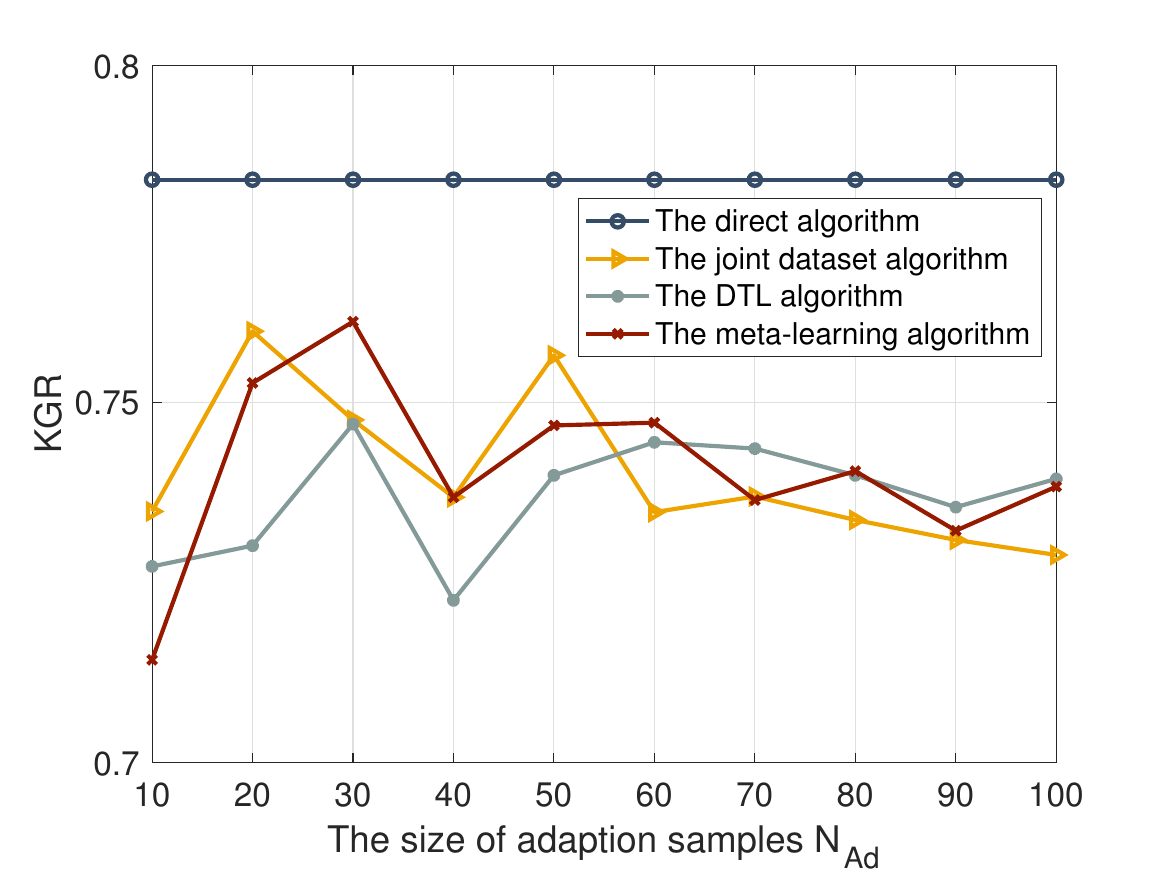}}  
\caption{The performance under different numbers of adaption samples.}
\label{fig_practical}     
\end{figure}

Next, we evaluate the performance of the proposed algorithms in multi-environments. In this paper, we test the algorithm proposed in this paper using 900 sets of data collected in the indoor scenario to improve the performance in the outdoor scenario when the amount of data is small. Fig.~\ref{fig_practical} compares the performance of several algorithms under different adaptation sample sizes. Fig.~\ref{fig:cdf-1} and Fig.~\ref{fig:cdf-2} show that the NMSE and KER performances of the algorithms both improve as the number of adapted samples increases. When the adaptive sample size is only 10, the NMSE performance of the algorithms proposed in this paper is poor, and it is not as good as the direct algorithm. However, after quantization, the mete-learning algorithm can still generate keys with lower KER when the adaption sample size is as small as 10 and 20. When the adaption sample size is 80, the KER of the key generated by the meta-learning algorithm is lower than 10\%. In addition, different from the simulation results, when the adaptive sample size exceeds 40, the KER of the key generated by the joint dataset algorithm is also smaller than that of the direct algorithm. This shows that in real-world environments, despite experiencing different scattering environments, different environments still have certain similarities, which can be used to improve performance in new environments. At this point, the performance of the two algorithms proposed in this paper is still better than the joint dataset algorithm.

Fig. \ref{fig:cdf-3} shows that the performance of KGR seems to be slightly degraded, and the KGR generated by all algorithms fluctuates between 0.7 and 0.8. This result is related to the quantization method we choose. The quantization based on the Gaussian distribution used in this paper may not perfectly fit the distribution of channel characteristics, thus causing KGR fluctuations. Since the fluctuation range is not large, it is quite normal and feasible to sacrifice a certain KGR to achieve a lower KER key. Since the fluctuation range is not large and the KER of the generated key is significantly reduced, in general, the DTL algorithm and the mata-learning algorithm can significantly improve the performance of the generated key in real-world scenarios.

\begin{table}[!t] 
\caption{NIST statistical test results of the experimental data.} 
	\centering
	\begin{tabular}{|p{3.5cm}<{\centering}| p{1.5cm}<{\centering} | p{1.5cm}<{\centering} |}
		\hline
		 \bfseries Test&\bfseries Indoor &\bfseries Outdoor\\ \hline
		 Approximate Entropy& 0.4391 &0.4405\\\hline
		 Block Frequency & 0.9559 & 0.3620\\\hline
		 Cumulative Sums  &0.2653 &0.5700\\\hline
		 Discrete Fourier Transform & 0.4913 &0.9087\\\hline
		 Frequency & 0.2159 &0.2159\\\hline
		 Ranking & 0.7745  &0.2395\\\hline
		 Runs & 0.5005 &0.5628\\\hline
		 \multirow{2}{*}{\shortstack{Serial }} & 0.4369 & 3622\\ &0.7237 & 0.4795 \\
		 \hline
	\end{tabular}
\label{table_randomness_2}
\end{table}

The NIST test is also used to test the randomness of the generated keys in the real world. Table \ref{table_randomness_2} shows that all cases for both indoor and outdoor environments pass the test with p-values much greater than 0.01.

\section{Conclusion}
\label{Conclusion}
In this paper, aiming at the problem of inapplicability of deep learning model caused by environmental changes, we formulated this problem as a learning-based problem, i.e., using knowledge from source environments to learn the feature mapping in the new environments, and proposed a DTL algorithm and a meta-learning algorithm to achieve fast key generation in multi-environment for FDD-OFDM systems. Simulation results showed that both algorithms can effectively improve the performance of generated keys in new environments. When the SNR=20 dB, the KERs of the keys generated by the DTL and meta-learning algorithms were reduced by 73.14\% and 77.3\%, respectively, compared with the method without adaptation (the direct algorithm) in the new environments. 
The complexity analysis showed that the meta-learning algorithm consumed less time than the DTL algorithm in the training stage, and these costs were acceptable in real-world applications. In addition, we built a USRP SDR-based testbed and verified the performance of the learning-based FDD-OFDM key generation method using real-world data for the first time. The results show that the proposed algorithm can significantly reduce the KER of generated keys, and only 80 samples in the new environment can reduce KER to 10\%.


\bibliographystyle{IEEEtran} 
\bibliography{IEEEabrv, myref}

\begin{thebibliography}{10}
\providecommand{\url}[1]{#1}
\csname url@samestyle\endcsname
\providecommand{\newblock}{\relax}
\providecommand{\bibinfo}[2]{#2}
\providecommand{\BIBentrySTDinterwordspacing}{\spaceskip=0pt\relax}
\providecommand{\BIBentryALTinterwordstretchfactor}{4}
\providecommand{\BIBentryALTinterwordspacing}{\spaceskip=\fontdimen2\font plus
\BIBentryALTinterwordstretchfactor\fontdimen3\font minus
  \fontdimen4\font\relax}
\providecommand{\BIBforeignlanguage}[2]{{%
\expandafter\ifx\csname l@#1\endcsname\relax
\typeout{** WARNING: IEEEtran.bst: No hyphenation pattern has been}%
\typeout{** loaded for the language `#1'. Using the pattern for}%
\typeout{** the default language instead.}%
\else
\language=\csname l@#1\endcsname
\fi
#2}}
\providecommand{\BIBdecl}{\relax}
\BIBdecl

\bibitem{zou2016survey}
Y.~Zou, J.~Zhu, X.~Wang, and L.~Hanzo, ``{A survey on wireless security:
  Technical challenges, recent advances, and future trends},'' \emph{Proc.
  IEEE}, vol. 104, no.~9, pp. 1727--1765, Sep. 2016.

\bibitem{li2019physical}
G.~Li, C.~Sun, J.~Zhang, E.~Jorswieck, B.~Xiao, and A.~Hu, ``{Physical layer
  key generation in 5G and beyond wireless communications: Challenges and
  opportunities},'' \emph{Entropy}, vol.~21, no.~5, p. 497, May 2019.

\bibitem{8735939}
J.~Zhang, M.~Ding, G.~Li, and A.~Marshall, ``{Key generation based on large
  scale fading},'' \emph{IEEE Trans. Veh. Technol.}, vol.~68, no.~8, pp.
  8222--8226, aug. 2019.

\bibitem{li2018constructing}
G.~Li, A.~Hu, C.~Sun, and J.~Zhang, ``{Constructing reciprocal channel
  coefficients for secret key generation in FDD systems},'' \emph{IEEE Commun.
  Lett.}, vol.~22, no.~12, pp. 2487--2490, Dec. 2018.

\bibitem{penttinen2015telecommunications}
J.~Penttinen, \emph{{The telecommunications handbook: Engineering guidelines
  for fixed, mobile and satellite systems}}.\hskip 1em plus 0.5em minus
  0.4em\relax Wiley, 2015.

\bibitem{3gpp.38.101-1}
\BIBentryALTinterwordspacing
3GPP, ``{NR; User Equipment (UE) radio transmission and reception; Part 1:
  Range 1 Standalone},'' Technical Specification (TS) 36.101-2, Jun. 2022,
  v17.6.0. [Online]. Available:
  \url{https://portal.3gpp.org/desktopmodules/Specifications/SpecificationDetails.aspx?specificationId=3283}
\BIBentrySTDinterwordspacing

\bibitem{wang2012wireless}
W.~Wang, H.~Jiang, X.~Xia, P.~Mu, and Q.~Yin, ``{A wireless secret key
  generation method based on Chinese remainder theorem in FDD systems},''
  \emph{Sci. China Inf. Sci.}, vol.~55, no.~7, pp. 1605--1616, Jul. 2012.

\bibitem{liu2019secret}
B.~Liu, A.~Hu, and G.~Li, ``{Secret key generation scheme based on the channel
  covariance matrix eigenvalues in FDD systems},'' \emph{IEEE Commun. Lett.},
  vol.~23, no.~9, pp. 1493--1496, Sep. 2019.

\bibitem{vasisht2016eliminating}
D.~Vasisht, S.~Kumar, H.~Rahul, and D.~Katabi, ``{Eliminating channel feedback
  in next-generation cellular networks},'' in \emph{Proc. ACM Conf. Special
  Interest Group Data Commun. (SIGCOMM)}, Florianopolis, Brazil, Aug. 2016, pp.
  398--411.

\bibitem{qin2016exploiting}
D.~Qin and Z.~Ding, ``{Exploiting multi-antenna non-reciprocal channels for
  shared secret key generation},'' \emph{IEEE Trans. Inf. Forensics Secur.},
  vol.~11, no.~12, pp. 2693--2705, Dec. 2016.

\bibitem{allam2017channel}
A.~M. Allam, ``{Channel-based secret key establishment for FDD wireless
  communication systems},'' \emph{Commun. Appl. Electron}, vol.~7, no.~9, pp.
  27--31, Nov. 2017.

\bibitem{linning2018investigation}
L.~Peng, G.~Li, J.~Zhang, R.~Woods, M.~Liu, and A.~Hu, ``{An investigation of
  using loop-back mechanism for channel reciprocity enhancement in secret key
  generation},'' \emph{IEEE Trans. Mob. Comput.}, vol.~18, no.~3, pp. 507--519,
  May 2019.

\bibitem{zhang2021secret}
X.~Zhang, G.~Li, Z.~Hou, and A.~Hu, ``{Secret key generation for FDD systems
  based on complex-valued neural network},'' in \emph{Proc. IEEE Veh Technol
  Conf (VTC2021-Fall)}, Virtual, Sep. 2021, pp. 1--6.

\bibitem{wan2021secret}
Z.~Wan, K.~Huang, and L.~Chen, ``{Secret key generation scheme based on deep
  learning in FDD MIMO systems},'' \emph{IEICE Trans Inf Syst}, vol. 104,
  no.~7, pp. 1058--1062, Jul. 2021.

\bibitem{10.1145/3522783.3529526}
X.~Wei and D.~Saha, ``{Knew: Key generation using neural networks from wireless
  channels},'' in \emph{Proc. ACM Workshop Wirel. Secur. Mach. Learn.
  (WiseML'22)}, San Antonio, TX, USA, May 2022, pp. 45--50.

\bibitem{hou2021secret}
Z.~Hou and X.~Zhang, ``{Secret key generation scheme based on generative
  adversarial networks in FDD systems},'' in \emph{Proc. IEEE Conf. Comput.
  Commun. Workshops (INFOCOM WKSHPS)}, Virtual, May 2021, pp. 1--6.

\bibitem{zhang2022deep}
X.~Zhang, G.~Li, J.~Zhang, A.~Hu, Z.~Hou, and B.~Xiao, ``{Deep-learning-based
  physical-layer secret key generation for FDD systems},'' \emph{IEEE Internet
  of Things J.}, vol.~9, no.~8, pp. 6081--6094, Apr. 2022.

\bibitem{alrabeian2019deep}
M.~Alrabeiah and A.~Alkhateeb, ``{Deep learning for TDD and FDD massive MIMO:
  Mapping channels in space and frequency},'' in \emph{Proc. Conf. Rec.
  Asilomar Conf. Signals Syst. Comput. (ACSSC)}, Pacific Grove, CA, United
  states, Nov. 2019, pp. 1465--1470.

\bibitem{bakshi2019fast}
A.~Bakshi, Y.~Mao, K.~Srinivasan, and S.~Parthasarathy, ``Fast and efficient
  cross band channel prediction using machine learning,'' in \emph{Proc. Annu
  Int Conf Mobile Comput Networking (Mobicom'19)}, Los Cabos, Mexico, Oct.
  2019, pp. 1--16.

\bibitem{liu2021Fire}
Z.~Liu, G.~Singh, C.~Xu, and D.~Vasisht, ``{FIRE: Enabling reciprocity for FDD
  MIMO systems},'' in \emph{Proc. Annu Int Conf Mobile Comput Networking
  (Mobicom'21)}, New Orleans, Louisiana, Oct. 2021, pp. 628--641.

\bibitem{nguyen2021transfer}
C.~T. Nguyen, N.~Van~Huynh, N.~H. Chu, Y.~M. Saputra, D.~T. Hoang, D.~N.
  Nguyen, Q.-V. Pham, D.~Niyato, E.~Dutkiewicz, and W.-J. Hwang, ``{Transfer
  learning for future wireless networks: A comprehensive survey},'' \emph{Proc.
  {IEEE}}, vol. 110, no.~8, pp. 1073--1115, Aug. 2022.

\bibitem{thrun1998learning}
S.~Thrun and L.~Pratt, ``{Learning to learn: Introduction and overview},'' in
  \emph{Learning to learn}.\hskip 1em plus 0.5em minus 0.4em\relax Springer,
  1998, pp. 3--17.

\bibitem{park2021learning}
S.~Park, H.~Jang, O.~Simeone, and J.~Kang, ``{Learning to demodulate from few
  pilots via offline and online meta-learning},'' \emph{IEEE Trans Signal
  Process}, vol.~69, pp. 226--239, Dec. 2021.

\bibitem{zeng2021downlink}
J.~Zeng, J.~Sun, G.~Gui, B.~Adebisi, T.~Ohtsuki, H.~Gacanin, and H.~Sari,
  ``{Downlink CSI feedback algorithm with deep transfer learning for FDD
  massive MIMO systems},'' \emph{{IEEE} Trans. on Cogn. Commun. Netw.}, vol.~7,
  no.~4, pp. 1253--1265, 2021.

\bibitem{yuan2020transfer}
Y.~Yuan, G.~Zheng, K.-K. Wong, B.~Ottersten, and Z.-Q. Luo, ``Transfer learning
  and meta learning-based fast downlink beamforming adaptation,'' \emph{{IEEE}
  Trans. Wireless Commun.}, vol.~20, no.~3, pp. 1742--1755, 2020.

\bibitem{yang2020deep}
Y.~Yang, F.~Gao, Z.~Zhong, B.~Ai, and A.~Alkhateeb, ``{Deep transfer
  learning-based downlink channel prediction for FDD massive MIMO systems},''
  \emph{IEEE Trans. Commun.}, vol.~68, no.~12, pp. 7485--7497, Dec. 2020.

\bibitem{9367008}
Y.~Ge and J.~Fan, ``{Beamforming optimization for intelligent reflecting
  surface assisted MISO: A deep transfer learning approach},'' \emph{IEEE
  Trans. Veh. Technol.}, vol.~70, no.~4, pp. 3902--3907, apr. 2021.

\bibitem{kingma2014adam}
D.~P. Kingma and J.~Ba, ``{Adam: A method for stochastic optimization},''
  \emph{arXiv:1412.6980}, 2014.

\bibitem{finn2017model}
C.~Finn, P.~Abbeel, and S.~Levine, ``Model-agnostic meta-learning for fast
  adaptation of deep networks,'' in \emph{Proc. Int. Conf. Mach. Learn.
  (ICML)}, Sydney, NSW, Australia, Aug. 2017, pp. 1126--1135.

\bibitem{massey1951kolmogorov}
F.~J. Massey~Jr, ``{The Kolmogorov-Smirnov test for goodness of fit},'' \emph{J
  AM STAT ASSOC}, vol.~46, no. 253, pp. 68--78, Mar. 1951.

\bibitem{terrell1992variable}
G.~R. Terrell and D.~W. Scott, ``{Variable kernel density estimation},''
  \emph{Ann. Stat.}, pp. 1236--1265, Sep. 1992.

\bibitem{richa2023sharing}
R.~Upadhyay, R.~Phlypo, R.~Saini, and M.~Liwicki, ``Sharing to learn and
  learning to share - fitting together meta-learning, multi-task learning, and
  transfer learning : {A} meta review,'' \emph{arXiv:2111.12146}, 2023.

\bibitem{Remcom}
\BIBentryALTinterwordspacing
``Remcom wireless insite.'' Sep. 2019. [Online]. Available:
  \url{http://www.remcom.com/wireless-insite}
\BIBentrySTDinterwordspacing

\bibitem{chen2023physical}
Y.~Chen, Z.~Chen, Y.~Zhang, Z.~Luo, Y.~Li, B.~Xing, B.~Guo, and L.~Chen,
  ``Physical layer key generation scheme for mimo system based on feature
  fusion autoencoder,'' \emph{IEEE Internet of Things J.}, 2023.

\bibitem{rukhin2001statistical}
A.~Rukhin, J.~Soto, J.~Nechvatal, M.~Smid, and E.~Barker, ``{A statistical test
  suite for random and pseudorandom number generators for cryptographic
  applications},'' Tech. Rep., 2001.

\bibitem{ettus}
\BIBentryALTinterwordspacing
``Ettus research,'' Jun. 2018. [Online]. Available: \url{http://www.ettus.com/}
\BIBentrySTDinterwordspacing

\end{thebibliography}

\begin{IEEEbiography}[{\includegraphics[width=1in,height=1.25in, clip,keepaspectratio]{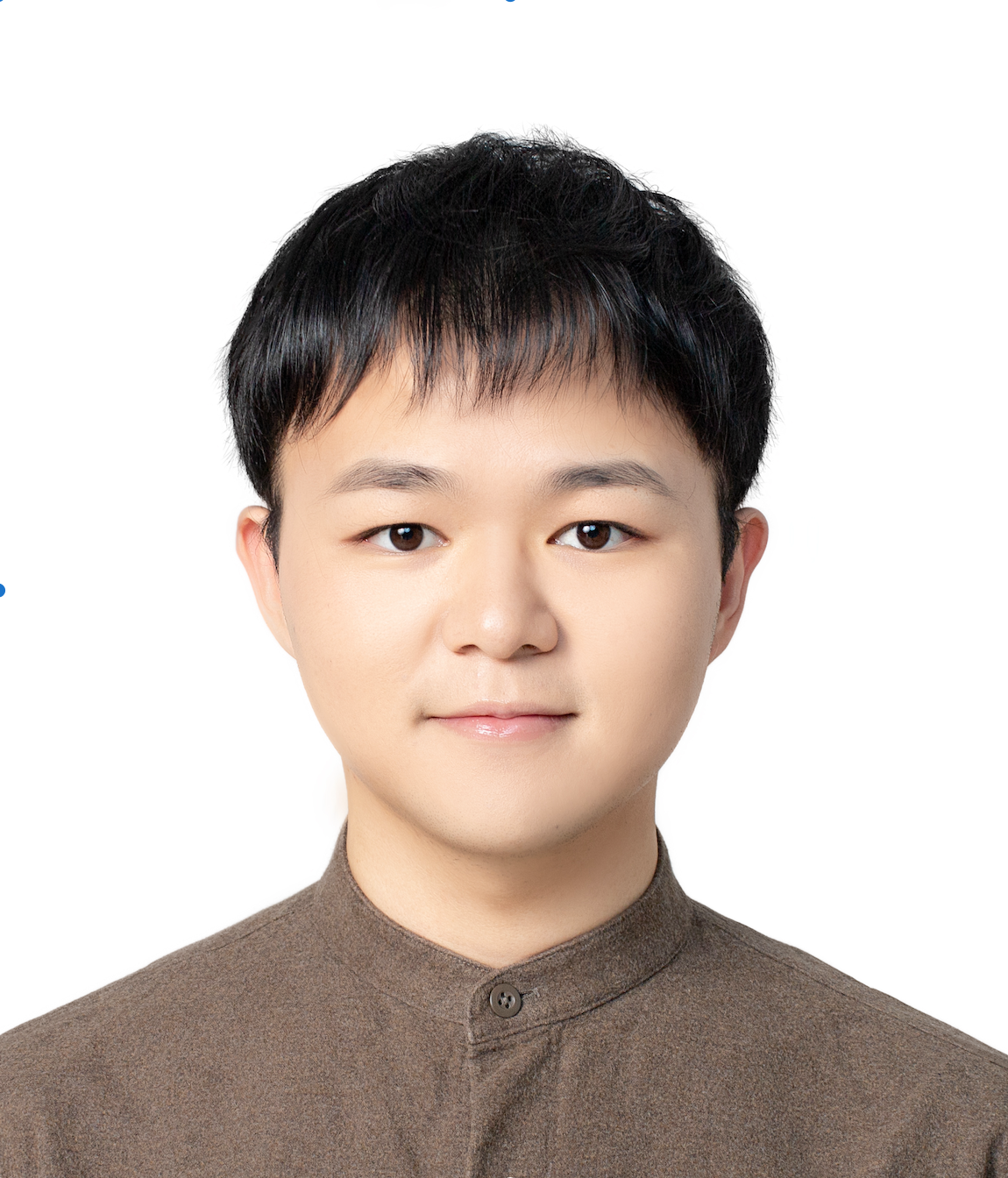}}]{Xinwei Zhang} received the M.Eng degree in computer technology from Southeast University, Nanjing, China, in 2022.  He is currently pursuing the Ph.D. degree with the Department of Electrical and Electronic Engineering, The Hong Kong Polytechnic University, Hong Kong. 

From April 2021 to September 2021, he was a Research Assistant with the Department of Computing, The Hong Kong Polytechnic University. His research interests include physical-layer security and adversarial machine learning.
\end{IEEEbiography}

\begin{IEEEbiography}[{\includegraphics[width=1in,height=1.25in,clip,keepaspectratio]{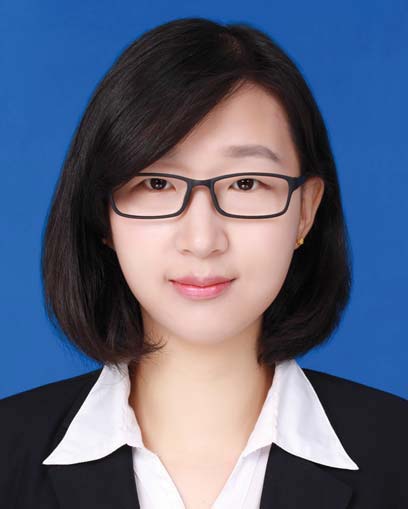}}]{Guyue Li}
(Member, IEEE) received the B.S. degree in information science and technology and the Ph.D. degree in information security from Southeast University, Nanjing, China, in 2011 and 2017, respectively. 

From June 2014 to August 2014, she was a Visiting Student with the Department of Electrical Engineering, Tampere University of Technology, Finland. She is currently an Associate Professor with the School of Cyber Science and Engineering, Southeast University. Her research interests include physical-layer security, secret key generation, radio frequency fingerprint, and link signature.
\end{IEEEbiography}

\begin{IEEEbiography}[{\includegraphics[width=1in,height=1.25in,clip,keepaspectratio]{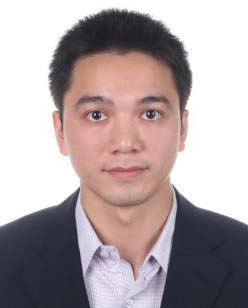}}]{Junqing Zhang}
(Member, IEEE) received the B.Eng and M.Eng degrees in Electrical Engineering from Tianjin University, China in 2009 and 2012, respectively, and the Ph.D degree in Electronics and Electrical Engineering from Queen's University Belfast, UK in 2016. From Feb. 2016 to Jan. 2018, he was a Postdoctoral Research Fellow at Queen's University Belfast. From Feb. 2018 to Oct. 2022, he was a Tenure Track Fellow and then a Lecturer (Assistant Professor) at the University of Liverpool, UK. Since Oct. 2022, he has been a Senior Lecturer (Associate Professor) with the University of Liverpool. His research interests include the Internet of Things, wireless security, physical layer security, key generation, radio frequency fingerprint identification, and wireless sensing. He was a recipient of the UK EPSRC New Investigator Award.
\end{IEEEbiography}

\begin{IEEEbiography}[{\includegraphics[width=1in,height=1.25in,clip,keepaspectratio]{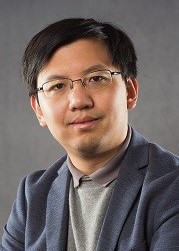}}]{Linning Peng}
(Member, IEEE) received the Ph.D.
degree from the Electronics and Telecommunications Institute of Rennes Laboratory, National Institute of Applied Sciences, Rennes, France, in 2014.

Since 2014, he has been an Associate Professor with Southeast University, Nanjing, China. His
research interests include Internet of Things and
physical layer security in wired and wireless communications

\end{IEEEbiography}

\begin{IEEEbiography}[{\includegraphics[width=1in,height=1.25in,clip,keepaspectratio]{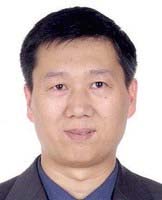}}]{Aiqun Hu}
(Senior Member, IEEE) received the B.Sc.(Eng.), M.Eng.Sc., and Ph.D. degrees from Southeast University in 1987, 1990, and 1993, respectively. 

He was invited as a Post-Doctoral Research Fellow with The University of Hong Kong from 1997 to 1998, and a TCT Fellow with Nanyang Technological University in 2006. He has published two books and more than 100 technical articles in wireless communications field. His research interests include data transmission and secure communication technology.
\end{IEEEbiography}

\begin{IEEEbiography}[{\includegraphics[width=1in,height=1.25in,clip,keepaspectratio]{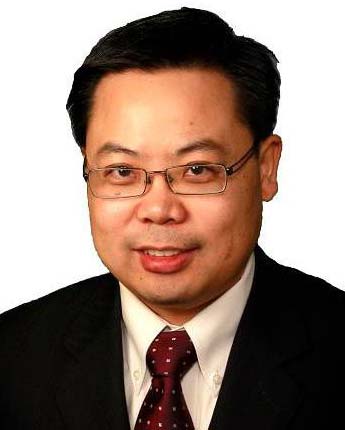}}]{Xianbin Wang} (Fellow, IEEE) received his Ph.D. degree in electrical and computer engineering from the National University of Singapore in 2001.

He is a Professor and a Tier-1 Canada Research Chair in 5G and Wireless IoT Communications with Western University, Canada. Prior to joining Western University, he was with the Communications Research Centre Canada as a Research Scientist/Senior Research Scientist from 2002 to 2007. From 2001 to 2002, he was a System Designer at STMicroelectronics. His current research interests include 5G/6G technologies, Internet of Things, machine learning, communications security, and intelligent communications. He has over 600 highly cited journals and conference papers, in addition to over 30 granted and pending patents and several standard contributions.

Dr. Wang is a Fellow of the Canadian Academy of Engineering and a Fellow of the Engineering Institute of Canada. He has received many prestigious awards and recognitions, including the IEEE Canada R. A. Fessenden Award, Canada Research Chair, Engineering Research Excellence Award at Western University, Canadian Federal Government Public Service Award, Ontario Early Researcher Award, and nine Best Paper Awards. He was involved in many IEEE conferences, including GLOBECOM, ICC, VTC, PIMRC, WCNC, CCECE, and CWIT, in different roles, such as General Chair, TPC Chair, Symposium Chair, Tutorial Instructor, Track Chair, Session Chair, and Keynote Speaker. He serves/has served as the Editor-in-Chief, Associate Editor-in-Chief, and editor/associate editor for over ten journals. He was the Chair of the IEEE ComSoc Signal Processing and Computing for Communications (SPCC) Technical Committee and is currently serving as the Central Area Chair of IEEE Canada.
\end{IEEEbiography}

\end{document}